\documentclass[preprint,12pt]{elsarticle}
\usepackage{booktabs}
\usepackage{makecell}
\usepackage{multirow}
\usepackage{longtable}
\usepackage{caption}
\usepackage{footnote}
\usepackage{graphicx}
\usepackage{float}
\usepackage{graphicx}
\usepackage{geometry}
\usepackage{url}
\usepackage{textcomp}
\usepackage{comment}
\usepackage{xurl}

\usepackage{amssymb}

\journal{arXiv}

\begin{document}

\begin{frontmatter}



\title{Zero Day Ransomware Detection with Pulse: Function Classification with Transformer Models and Assembly Language}

\author[inst1]{Matthew Gaber}

\affiliation[inst1]{organization={Edith Cowan University School of Science},
            addressline={270 Joondalup Dr}, 
            city={Joondalup },
            postcode={6027}, 
            state={WA},
            country={Australia}}

\author[inst1]{Mohiuddin Ahmed}
\author[inst1]{Helge Janicke}

\begin{abstract}
Finding automated AI techniques to proactively defend against malware has become increasingly critical. The ability of an AI model to correctly classify novel malware is dependent on the quality of the features it is trained with and the authenticity of the features is dependent on the analysis tool. Peekaboo, a Dynamic Binary Instrumentation tool defeats evasive malware to capture its genuine behavior. The ransomware Assembly instructions captured by Peekaboo, follow Zipf's law, a principle also observed in natural languages, indicating Transformer models are particularly well-suited to binary classification. We propose Pulse, a novel framework for zero day ransomware detection with Transformer models and Assembly language. Pulse, trained with the Peekaboo ransomware and benign software data, uniquely identify truly new samples with high accuracy. Pulse eliminates any familiar functionality across the test and training samples, forcing the Transformer model to detect malicious behavior based solely on context and novel Assembly instruction combinations.

\end{abstract}


\end{frontmatter}


\section{Introduction}
\label{sec:intro}
Due to the rapid increase in the use of digital devices and widespread internet access, the frequency of cyber-attacks targeting information systems is escalating. Malicious software (malware) is the primary weapon used by attackers in executing these cyber-attacks. Various types of malware such as ransomware, trojans, botnets and spyware are continually evolving, incorporating advanced encryption and obfuscation techniques \cite{gabercsur2024}. Further, each type of malware has distinct objectives and functionality \cite{xiao2020}. This research focuses on developing Pulse, a framework that uses Transformer models and the Peekaboo Dynamic Binary Instrumentation (DBI) data to detect zero day ransomware \cite{gaber2024}.

According to Sophos, 59\% of 5000 surveyed organizations across 14 countries were hit by ransomware in 2023, where the average ransom demand was USD 4,321,880 \cite{sophos2024}. Further, IBM report the average cost of a data breach was USD 4.45 million in 2023, while the average time to identify and contain a breach was 277 days \cite{ibmsecurity2023}. Ransomware is designed to encrypt victims' files \cite{kerns2022, payne2021}. Recently, ransomware gangs have added data theft to their tactics, targeting sensitive information including personal details and intellectual property. They threaten to publish or sell this data if the ransom isn't paid, a method known as double extortion \cite{kerns2022, payne2021}. This approach often results in larger ransoms and higher payments, boosting profits compared to attacks involving encryption alone \cite{meurs2022}. 

Software and malware consist of sets of data, instructions and functions designed for specific tasks, with many distributed in an executable binary format \cite{ahn2022}. Anti-malware vendors commonly employ signature-based methods to identify known threats, and heuristic based detection \cite{aurangzeb2021, HIRANO2022, khan2020}. A signature refers to a sequence of bytes derived from unique data within a binary \cite{khan2020, CARLIN2019, gibert2021}. When new malware emerges, vendors obtain a sample and define its signature based on strings, URLs, code segments and other patterns in the binary \cite{khan2020, CARLIN2019, gibert2021}. However, this process is manual and time-consuming, involving both the creation of the signature and its distribution through updates \cite{khan2020}. Moreover, recycled malware that has been altered or obfuscated can evade signature-based detection systems \cite{gibert2021}. Creating signatures for new or modified malware is time-consuming due to the analysis required and the client update, which may not occur for months. \cite{ye2017, khan2020, CARLIN2019}. Further,  with the rapid growth of AI, it is likely that malware will soon use AI techniques, specifically ransomware could adapt encryption methods intelligently and dynamically, making it harder to detect \cite{vonderassen2023}. Consequently, finding automated AI techniques to proactively defend against malware has become increasingly critical. However, the ability of an AI model to correctly classify novel malware is dependent on the quality of the data and features it is trained with. Further the authenticity and veracity of the features is dependent on the analysis tool and the dataset \cite{gabercsur2024, Kajiwara2021, ucci2019}.

Evasive malware easily defeats widely used static and dynamic analysis tools due to their various limitations, whereas DBI facilitates the capture of genuine behaviour from sophisticated malware \cite{galloro2022,kim2022,maffia2021,park2019,gaber2024}. If behavioural data is extracted from malware while it is concealing its malicious intent and true behaviour in an analysis environment, any AI model that is subsequently trained with that data would not correctly identify the malware when it executed in a non-analysis environment and revealed it malicious intent \cite{galloro2022, nunes2022}. Therefore, the data captured by Peekaboo was used in this research \cite{gaberro2024}. Peekaboo is an automated DBI tool that defeats 97 evasive techniques variously used by malware, then captures the malicious behaviour of the malware by executing each sample and recording every Assembly (ASM) instruction that is executed by the CPU \cite{gaber2024}. Peekaboo outperforms static and dynamic analysis tools by forcing the malware to reveal its malicious intent and then extracting it's authentic behavior \cite{gaber2024}.

Research employing ASM instructions with AI models has focused on Binary Code Similarity Detection (BCSD) tasks \cite{ahn2022,koo2021}. BCSD refers to the task of identifying similarities between different binary files based on ASM language rather than their source code. This is particularly important in the fields of software analysis where understanding the similarity between binary files can help in analysing software evolution and detecting code plagiarism \cite{ahn2022,koo2021}. Understanding the contextual information embedded in binary files presents significant challenges for several reasons. First, the structure and content of binaries, which are semantically equivalent, can differ depending on factors like obfuscation methods, compilers, optimizations, and architectures. Second, the complex compilation process often strips away semantic details, such as variable names, data structures, types, and class hierarchies. Finally, demonstrating the functional equivalence of any two arbitrary programs is extremely challenging \cite{ahn2022}. Similarly, sophisticated and novel malware detection techniques could leverage contextual meanings extracted from binary code, enabling the inference of code semantics from syntactically distinct samples. 

Fundamentally, the challenge is deducing previously unseen malicious behaviours based on contextual information and code semantics. That is, can an AI model effectively classify functions it encounters for the first time, as benign or malicious.

Considering the limitations of novel malware detection, the central question this research seeks to answer is how authentic behavioural data can be used with a Transformer model to detect novel malicious functionality. 

\subsection{Key Contributions}
Our main contributions are summarized as follows:
\begin{itemize}
    \item We developed Pulse, a framework that utilizes Peekaboo DBI ransomware data with Transformer models. \item Pulse uses a feature engineering approach that extracts contextual meaning and code semantics from the Peekaboo ASM language data.
    \item The experimental results demonstrate the effectiveness and transferability of Pulse, which is robust to detecting new malicious functions. 
    \item Pulse outperforms the previous state-of-the-art approaches for zero day ransomware detection.
    \item To the best of our knowledge using ASM language with Transformer models to detect new malicious functions and samples has not been covered by previous works. In the spirit of open science, we release the fine tuned Transformer models and associated scripts \cite{gabertpro2024}.
\end{itemize}

\subsection{Paper Roadmap}
The remainder of this paper is organized as follows. Section 2 presents background information on the fundamental concepts used in this research. Section 3 describes the related works. Section 4 presents the proposed ransomware detection pipeline in detail. Section 5 describes the experimental analysis and details the implementation. Section 6 discusses the results. Section 7 presents the conclusions and future research.

\section{Background}
\label{sec:background}
In this section, we provide background information on Transformer models, the Peekaboo DBI data and Zipfs law.
\subsection{Transformer models}
Transformer models have revolutionized Natural Language Processing (NLP) by providing efficient ways to process and generate text, leveraging self-attention mechanisms to capture dependencies and contextual information effectively across long range sequences \cite{devlin2019, vaswani2017}. The Transformer model does not use recurrence or convolutions, only attention mechanisms \cite{vaswani2017}. There are several key components of a Transformer model. The Attention Mechanism allows the model to relate the importance of different parts of the input sequence when generating a representation of the sequence \cite {vaswani2017}. This is achieved by calculating attention scores between all pairs of positions in the input sequence. To account for the sequential nature of the input data, Positional Encodings that convey information about the order of tokens in the sequence are added to the input \cite {vaswani2017}. The Encoder processes the input sequence through multiple layers of self-attention mechanisms and feed forward neural networks. Each layer in the encoder refines the representation of the input sequence by capturing dependencies between tokens (words) at different positions \cite {vaswani2017}. The Decoder also consists of multiple layers, incorporating self-attention mechanisms and feed forward neural networks. The Decoder generates the output sequence based on the encoder's final hidden representations \cite {vaswani2017}. However, the Encoder is not typically used for classification tasks; instead, it generates outputs such as text. For classification tasks, a Classification Head is added on top of the Encoder's output. It is normally a simple feed forward network, that learns to classify inputs by optimizing the Classification Head along with the Encoder layers. During inference, the model uses the Classification Head to predict the class label based on the encoded representation of the input sequence \cite{devlin2019}.

Bidirectional Encoder Representations from Transformers (BERT) are used for language modelling, text classification, and question answering, where understanding context and semantics are crucial. Unlike earlier models that process sequences in one direction, BERT captures bi-directional context \cite{devlin2019}. BERT is initially pre-trained using large amounts of unlabelled text, where the objective is to learn general language representations that capture semantic relationships and context across different sentences. BERT introduced the Masked Language Model (MLM) pre-training objective, where it masks random words in sentences and trains the model to predict these masked words based on surrounding context, this helps BERT learn deeper semantic relationships by understanding the bidirectional context and dependencies within sentences \cite{devlin2019}. BERT is also trained on a classification task, that is Next Sentence Prediction (NSP), where it predicts whether two input sentences follow each other logically in a given document which helps BERT understand relationships between sentences and improves its ability to generate coherent outputs \cite{devlin2019}.

DistilBERT differs from the original BERT model primarily in terms of size and computational efficiency while aiming to preserve as much of BERT's performance as possible \cite{sanh2020}. DistilBERT uses parameter sharing techniques to reduce the number of parameters and is trained using a combination of supervised learning and distillation from a larger, pre-trained BERT model \cite{sanh2020}.

Robustly optimized BERT approach (RoBERTa) retains the same base Transformer architecture as BERT but focuses on enhancing training methodology and data scale \cite{liu2019}. RoBERTa eliminates the NSP task and performs extensive hyperparameter tuning compared to BERT  \cite{liu2019}. BERT was shown to be under trained, RoBERTa is trained for a longer period of time on 160GB data, as compared to 16GB for BERT \cite{liu2019}. DistilRoBERTa is a distilled version of RoBERTa and follows the same training procedure as DistilBERT \cite{huggingface2024}.

Generative Pre-trained Transformer 2 (GPT-2) uses a Transformer architecture that consists of just the decoder stack \cite{radford2019}. GPT-2 is trained using an autoregressive language modelling objective and it generates text sequentially, predicting the next word in a sequence based on preceding words \cite{radford2019}. Unlike BERT, which utilizes bidirectional context, GPT-2 employs a left-to-right language modeling approach. GPT-2 is primarily designed for generative tasks such as text generation, story completion, and question answering \cite{radford2019}.

The eXtreme Language understanding Network (XLNet) model uses an autoregressive pretraining method that considers all permutations of the input sequence using Permutation Language Modelling (PLM) which allows it to learn bidirectional context similar to BERT but without explicitly masking tokens \cite{yang2019}. XLNet's training that uses bidirectional context and dependencies makes it particularly effective for tasks that require deep understanding of relationships within and across sentences and documents \cite{yang2019}.

\begin{table}[!htb]
\centering
\footnotesize
\captionsetup{justification=centering}
  \caption{Binary corpus of ransomware and benign samples from Peekaboo.}
  \label{tab:binarycorpus}
  \begin{tabular}{ccccc}
    \toprule
     \makecell{Family} & \makecell{Samples} & \makecell{Total\\Instructions} & \makecell{Unique\\Instructions} & \makecell{Unique\\Instructions\\Freq. \textgreater{}10}\\
    \midrule
       \makecell{BlackCat} & \makecell{62} & \makecell{27,002,671} & \makecell{2,619,503} & \makecell{47,509} \\
        \makecell{Chaos} & \makecell{2} & \makecell{3,184,930} & \makecell{309,300} & \makecell{19,303} \\
        \makecell{Clop} & \makecell{2} & \makecell{622,184} & \makecell{68,003} & \makecell{5,778} \\
        \makecell{Conti} & \makecell{51} & \makecell{24,999,412} & \makecell{378,310} & \makecell{118,402} \\
        \makecell{DarkSide} & \makecell{47} & \makecell{18,966,361} & \makecell{158,799} & \makecell{77,016} \\
        \makecell{Dharma} & \makecell{10} & \makecell{5,273,196} & \makecell{155,668} & \makecell{56,320} \\
        \makecell{Hive} & \makecell{15} & \makecell{2,945,110} & \makecell{180,957} & \makecell{27,402} \\
        \makecell{LockBit} & \makecell{26} & \makecell{11,971,432} & \makecell{219,735} & \makecell{82,199} \\
        \makecell{Locky} & \makecell{17} & \makecell{5,731,754} & \makecell{245,030} & \makecell{45,811} \\
        \makecell{Maze} & \makecell{9} & \makecell{7,652,149} & \makecell{203,378} & \makecell{64,308} \\
        \makecell{NetWalker} & \makecell{33} & \makecell{11,036,311} & \makecell{171,614} & \makecell{73,423} \\
        \makecell{Petya} & \makecell{5} & \makecell{2,183,958} & \makecell{165,685} & \makecell{19,859} \\
        \makecell{RagnarLocker} & \makecell{2} & \makecell{4,426,541} & \makecell{323,129} & \makecell{29,165} \\
        \makecell{Ryuk} & \makecell{26} & \makecell{6,010,623} & \makecell{282,894} & \makecell{38,511} \\
        \makecell{Sodinokibi} & \makecell{70} & \makecell{60,355,449} & \makecell{378,219} & \makecell{154,552} \\
        \makecell{Stop} & \makecell{989} & \makecell{203,136,528} & \makecell{601,964} & \makecell{98,197} \\
        \makecell{WannaCry} & \makecell{48} & \makecell{24,750,278} & \makecell{217,701} & \makecell{115,417} \\
        \makecell{Benign} & \makecell{340} & \makecell{80,081,404} & \makecell{1,735,692} & \makecell{196,337} \\
        \makecell{Benign-D} & \makecell{317} & \makecell{73,663,725} & \makecell{1,579,608} & \makecell{184,621} \\
    
    \bottomrule
\end{tabular}
\end{table}

\subsection{Peekaboo DBI}
The Peekaboo dataset consists of authentic behavioural data from 18,527 malware samples across ransomware, botnets, spyware, worms, trojans, Advanced Persistent Threats (APT) and post exploitation tools where every sample includes type, family, and variant information, for example Ransomware-WannaCry-SHA256. There are also 1,973 benign software samples \cite{gaber2024}.

In this research we used the Peekaboo ransomware and benign data which consists of 2,071 binaries with 573,994,016 ASM instructions and operands \cite{gaberro2024}. The unique ASM instruction counts range from BlackCat at 2,619,503 to Clop at 68,003, as shown in Table \ref{tab:binarycorpus}. Instructions such as \texttt{push esi} are used thousands of times, which explains the large difference between total and unique instructions.

\begin{figure}[!htb]
  \centering
  \includegraphics[width=\textwidth]{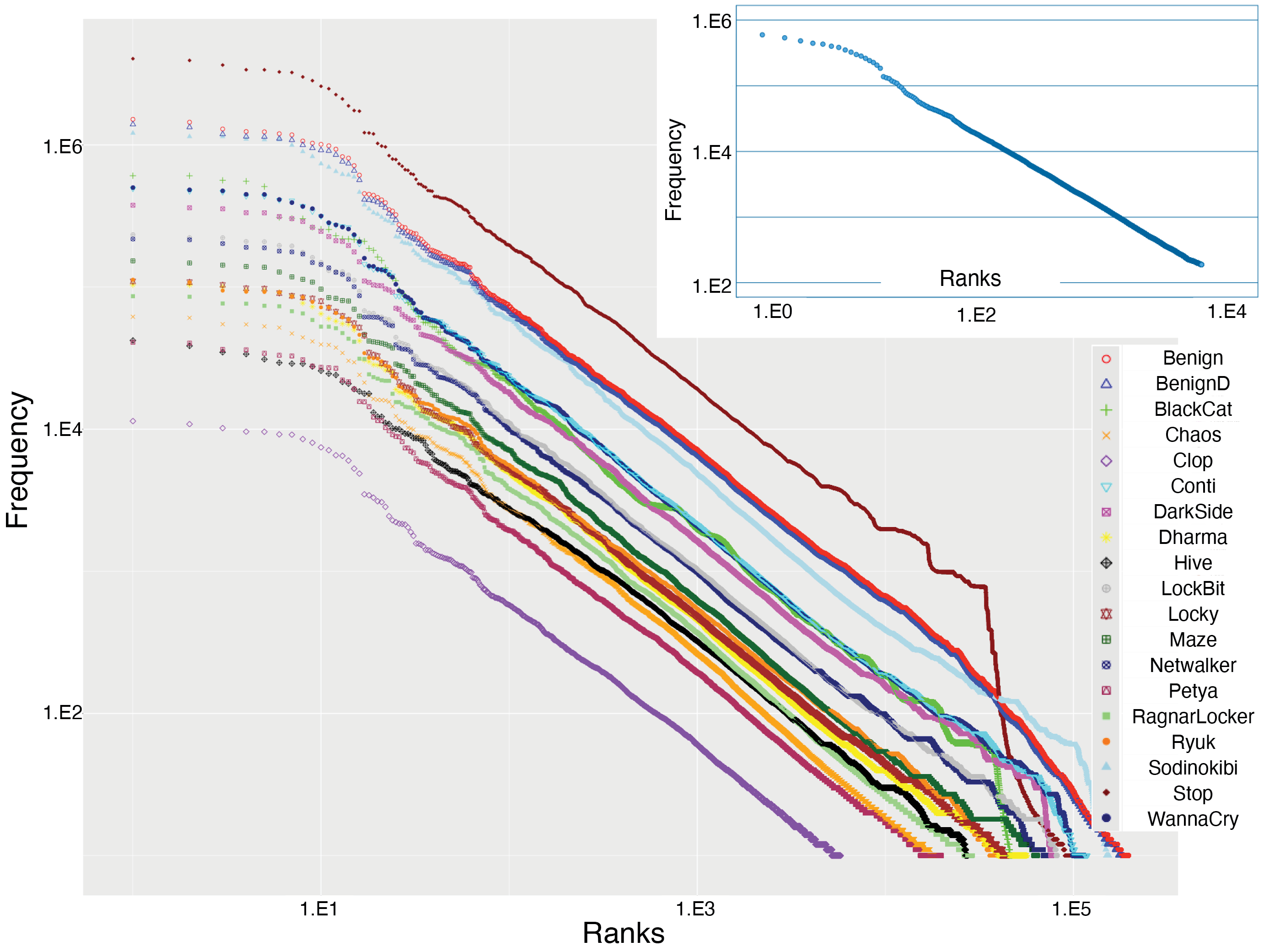}
  \caption{Log log plot of the frequency of each ASM instruction as a function of its frequency rank used by the Peekaboo ransomware and benign samples, inset shows the ideal with an alpha=0.1.}
  \label{fig:zipfcombined}
\end{figure}

\subsection{Zipfs law}
Assembly language displays linguistic traits that provide insights into its structural composition and patterns of utilization. In natural languages, word frequencies approximately follow the Zipf distribution that describes the inverse relationship between word frequency and word rank \cite{thurner2015}. The rank ordered distribution of word frequencies is described by the following approximate power law \cite{thurner2015}.

\begin{equation}
f(r) \sim r^a
\end{equation}

The frequency rank for the Peekaboo ransomware and benign ASM instructions is shown in Figure \ref{fig:zipfcombined}. The inset shows the ideal Zipfs law with an a=0.1. This indicates a skewed distribution where the frequencies of ASM instructions decrease slowly as their ranks increase. The inverse relationship when plotted on log scales predict a straight line, which is observed for all samples in Figure \ref{fig:zipfcombined} and indicates the ASM instructions used in the samples follows Zipfs law.

The ASM language used by the samples closely follows Zipf's law indicating Transformer models would be effective for binary classification \cite{koo2021}. Transformer models are pre-trained on large amounts of text data and are capable of capturing contextual information. In languages that adhere to Zipf's law, frequently used words such as \texttt{the}, \texttt{of}, and \texttt{and} or ASM instructions such as \texttt{mov}, \texttt{push} and \texttt{pop} appear very frequently and are crucial for understanding the context of words and sentences, or in this case, instructions and functions. The ability of Transformer models to capture context helps in accurately predicting the meaning of less frequent words based on their context in the sentence \cite{devlin2019, vaswani2017,radford2019}. That is, Transformer models excel at understanding and representing the meaning of low frequency words by leveraging their context within sentences. Transformer models generate word embeddings that not only capture syntactic relationships between words but also semantic relationships. This means that even if two words rarely co-occur, the models can still infer similarities or relationships between them based on their embeddings learned during pre-training. Consequently, several Transformer models were fine-tuned and tested using the ASM instructions from Peekaboo. The underlying idea is that ASM instructions and their operands function similarly to words, whereas functions are analogous to sentences.

\section{Related Works}
\label{sec:relwork}
Research studies in this field that have used ASM instructions with Transformer models have focused on BCSD tasks, such as identifying vulnerabilities and detecting software plagiarism. Further, API call sequences and features derived from static analysis have been used with Transformer models for malware detection. These tasks necessitate extracting contextual meanings from binary code. This section provides a summary of recent research and their focus, where
Table \ref{tab:litreviews} summarizes the contributions. 

PalmTree \cite{li2021} used an encoding technique with a small BERT assembly language model and assessed its performance by benchmarking it against other encoding techniques such as Instruction2Vec, word2vec, and Asm2Vec. Intrinsic evaluations covered outlier detection and basic block search. PalmTree consistently outperformed baseline schemes in both tasks, demonstrating its ability to effectively capture semantic differences in opcodes and operands, highlighting its capability to model assembly language effectively. Extrinsic evaluation assesses how well an instruction embedding model serves as input for downstream learning tasks, focusing on specific applications in binary analysis. Three key tasks were chosen for evaluation, BCSD, Function Type Signature Analysis (FTSA), and Value Set Analysis (VSA). Despite significant differences between training and testing datasets, PalmTree consistently demonstrates strong performance in BCSD and outperforms other schemes, indicating its enhanced generalisation across diverse datasets and compilers \cite{li2021}. FTSA refers to the process of identifying and categorizing the types of functions based on their signatures or prototypes within a software system. A function signature typically includes information such as the function's name, return type, and parameter types. PalmTree outperformed baselines, including Instruction2Vec and Asm2Vec, highlighting its ability to accurately represent FTSA across diverse datasets generated by different compilers. \cite{li2021}. In practical terms, VSA helps in understanding and predicting how data is accessed and manipulated within a program, which is crucial for tasks such as optimizing code performance, detecting memory-related bugs, understanding program behavior, and improving software security. PalmTree consistently outperformed the original DeepVSA and other baseline models, particularly excelling in categorizing global and heap memory references \cite{li2021}.

BinShot was presented by \cite{ahn2022}, a BERT based Siamese architecture for BCSD. IDAPro was used to statically disassemble the binary files, which did not use obfuscation, and convert the ASM into normalised functions using the rules proposed by DeepSemantic \cite{koo2021}. This approach could not be used for malware, which is heavily and widely obfuscated \cite{gibert2021,khan2020,xiao2020}. Further, after the normalization process two different functions may be mapped into the same normalized function and they were excluded from the training data \cite{ahn2022}. NSP is excluded from their BERT model as function invocations define the relationship between functions in binary code, not the sequential order \cite{ahn2022}. BinShot uses a BERT model to embed the normalised functions and uses a Siamese neural network to compute distances between function pairs based on these embeddings, and trains a binary classifier to predict function pair similarity using a weighted distance vector and binary cross-entropy loss \cite{ahn2022}. BinShot outperforms other BCSD models, including Gemini, Asm2Vec, PalmTree and DeepSemantic in predicting the similarity between two functions \cite{ahn2022}. However, it should be noted that after the function normalization the binary corpus contained 18,449 tokens.

DeepSemantic is an instruction normalisation process with pre-trained and fine-tuned BERT models for BCSD \cite{koo2021}. An ASM language instruction normalization process is essential for use with neural networks \cite{koo2021}. Prior methods varied widely in how they simplified instructions, with some stripping away too much detail like immediate values, while others broke down instructions into excessively granular tokens, which can lead to difficulties in embedding, understanding semantics and Out Of Vocabulary (OOV) problems \cite{koo2021}. The proposed solution seeks a middle ground by normalizing instructions to accurately reflect their semantics while managing the number of distinct tokens. This approach ensures that important details such as immediate values, that can represent jump targets or call destinations, and register sizes are preserved appropriately. It also maintains the integrity of memory access patterns and pointer expressions \cite{koo2021}.

Various deep learning models for highly imbalanced multiclass malware classification based on API calls have been evaluated \cite{demirkiran2022}. The models were evaluated using AUC and F1-scores due to the imbalance in the four datasets that were used. The main findings revealed that a Transformer model with one transformer block achieved slightly better results than the Long Short Term Memory (LSTM) model \cite{demirkiran2022}. Additionally, pre-trained BERT and CANINE models outperformed the one-block Transformer architecture. However, Transformer and LSTM models were noticeably faster in both training and inference compared to BERT and CANINE. Despite this, the proposed bagging-based random transformer forest (RTF) model, an ensemble of BERT or CANINE models, proved to be more practical, considering that training time does not directly impact response time and the differences in inference time were minimal \cite{demirkiran2022}.

MalBERT is a model built on BERT that conducts static analysis of Android application source code. It uses pre-processed features to identify and classify malware into various categories \cite{rahali2021}. The Manifest.xml file was processed and a text based feature representation was used \cite{rahali2021}. The BERT based model achieved a high accuracy, surpassing other baseline pre-trained language models in both binary and cross-category classification tasks \cite{rahali2021}. The experimental results showed a peak binary accuracy of 97.61\% and a multi-class accuracy of 91.02\% \cite{rahali2021}. This indicates that using Transformer-based models with text input for feature representation is highly effective. Further, pre-trained Transformer models can outperform traditional Recurrent Neural Network (RNN models), such as LSTM, in cyber security applications, especially when applied to datasets like Androidzoo \cite{rahali2021}.

EarlyMalDetect utilizes a fine-tuned GPT-2 model that analyzes API calls to predict the subsequent API functions for the early detection of malware attacks \cite{maniriho2024}. The proposed approach focuses on sequence prediction using Transformer models, bidirectional Gated Recurrent Unit (GRU), and fully connected neural networks. The results show the model is highly effective in detecting and classifying unknown Windows malware \cite{maniriho2024}. However, the authors note that the dataset was small and the fine-tuned model can generate irrelevant sequences \cite{maniriho2024}.

SeMalBERT is a semantic based model that uses API call sequences and BERT, Convolutional Neural Network (CNN) and LSTM deep learning models \cite{LIU2024SeMalBERT}. Static analysis was used to extract API function call sequences from the malware binary files. The BERT model then pre-processes these sequences, accounting for semantic order and differences in API function calls across various malware types. The output from the BERT model are subsequently used as input to train the CNN LSTM model for malware detection. The results demonstrate that SeMalBERT is less affected by transformations in the malware achieving an accuracy of 98.81\%, that surpasses previous API function call approaches \cite{maniriho2024}. 

MALSIGHT is a framework designed for generating summaries of binary malware by analyzing both malicious source code and benign pseudocode \cite{lu2024malsightexploringmalicioussource}. Two types of summaries are constructed, MalS and MalP, using a Transformer model and refining them through manual interaction. The core of the framework, MalT5, is a Transformer-based code model trained on these datasets. During the testing phase, MalT5 iteratively processes pseudocode functions to produce detailed summaries, enhancing understanding of code structure and function interactions \cite{lu2024malsightexploringmalicioussource}. MalT5, with 0.77 billion parameters, achieves performance comparable to larger models like ChatGPT3.5, demonstrating the framework's efficiency and effectiveness \cite{lu2024malsightexploringmalicioussource}.

\begin{table}
\centering
\captionsetup{justification=centering}
  \caption{Summary and comparison of recent Transformer model research with our paper}
  \label{tab:litreviews}
  \begin{tabular}{ccccc}
    \toprule
     \makecell{Paper} & \makecell{Dynamic\\Analysis}&  \makecell{Assembly\\Language}& \makecell{Function\\Level}& \makecell{Malware or\\Ransomware\\Classification} \\
    \midrule
    \cite{li2021}&\texttimes&\checkmark&\checkmark&\texttimes\\
    \cite{ahn2022}&\texttimes&\checkmark&\checkmark&\texttimes\\
    \cite{koo2021}&\texttimes&\checkmark&\checkmark&\texttimes\\
    \cite{demirkiran2022}&\texttimes&\texttimes&\texttimes&\checkmark\\
    \cite{rahali2021}&\texttimes&\texttimes&\texttimes&\checkmark\\
    \cite{maniriho2024}&\texttimes&\texttimes&\texttimes&\checkmark\\
    \cite{LIU2024SeMalBERT}&\texttimes&\texttimes&\texttimes&\checkmark\\
    \cite{lu2024malsightexploringmalicioussource}&\texttimes&\texttimes&\checkmark&\checkmark\\
    This paper&\checkmark&\checkmark&\checkmark&\checkmark\\
    \bottomrule
\end{tabular}
\end{table}

\section{Proposed ransomware detection pipeline}
\label{sec:methods}
BERT, RoBERTa, XLNet and GPT-2 base transformer models were fine-tuned with labeled Peekaboo ransomware and benign data, as detailed in Table \ref{tab:binarycorpus}, and evaluated on their ability to classify never before seen functionality as malicious or benign. Fine-tuning adapts the pre-trained BERT model to perform more effectively on specific tasks by adjusting its parameters through additional training on task-specific datasets \cite{devlin2019}.
\begin{figure}[H]
  \centering
  \includegraphics[width=0.8\textwidth]{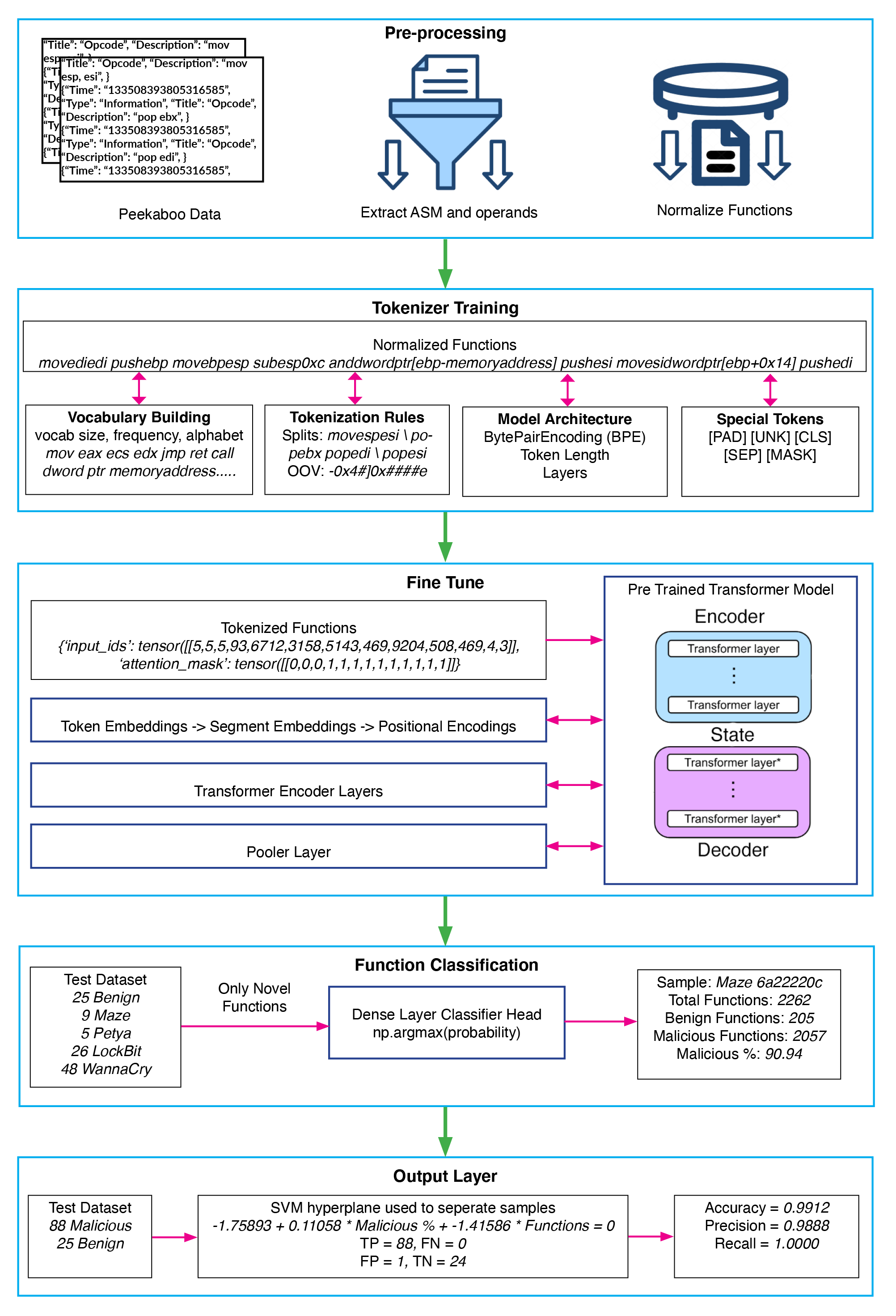}
  \caption{Pulse architecture}
  \label{fig:System design of Peekaboo Classifier}
\end{figure}

 A simple feedforward neural network and classification layer were appended to the pre-trained BERT model, and the combined system was trained end-to-end using the labeled Peekaboo data. The classification layer uses a softmax activation function and this provides probabilities for each class, enabling the model to make predictions. Softmax converts the output into probabilities, where each output node represents the probability of the input belonging to a particular class. During prediction the model takes new input data, processes them through the layers, and outputs a probability distribution over the two classes, that is benign or malicious. The class with the highest probability is chosen as the predicted class for the input.
 
GPT-2 is primarily designed for generative tasks such as text generation, story completion, and question answering \cite{radford2019}. A linear layer was used to transform the output from the GPT-2 model into logits. Logits are unnormalized scores assigned to each class and reflect the model's confidence in the predictions. They are computed directly from the linear transformation of the embeddings and are not scaled to a probability distribution yet. This step is pivotal in the classification process because it allows the model to learn discriminative patterns from the input data. During prediction, these logits are processed further with an argmax function, to obtain probabilities across the two classes. The class with the highest probability is then chosen as the predicted class by the model.

The ASM instructions captured by Peekaboo for both the malware and benign samples follows Zipf’s law which describes word frequency distribution in natural languages. Consequently, our approach involved extracting the ASM instructions and operands from the Peekaboo reports and variously engineering the data into the equivalent of words and sentences and then training custom tokenizers, as detailed in Tables \ref{tab:ASM Instruction Transformations}, \ref{tab:Function definitions}, \ref{tab:tokenizer details}. In this way an ASM instruction corresponds to a word and functions are equivalent to sentences. Every sample in the training dataset was processed into its constituent functions that were labelled accordingly, that is benign or malicious. At this granular level, where we examine ASM instructions, the ransomware samples contain functions that are also found in the benign samples. Consequently, any function present in the malware samples that also appeared in the benign samples was excluded from the malicious dataset.

Various Transformer models were trained with the different feature engineering approaches. The test data consisted of several complete ransomware families and benign samples that were not used in training. Given the main objective of this research was novel ransomware detection, any function present in the test samples that also appeared in the training data was removed from the test samples. Consequently, for the model to make the correct classification it had to infer what ASM instruction combinations and context were malicious and which were benign. The classification layer predicted the number of benign functions and malicious functions for each sample. The output layer performed the final classification for the test samples by utilizing a Support Vector Machine (SVM) hyperplane, which acts as the decision boundary to delineate between benign and malicious samples, as detailed in Figure \ref{fig:System design of Peekaboo Classifier}.

\subsection{Feature Engineering}
One of the challenges of feeding ASM language directly into a Transformer model is the vast number of possible tokens due to varying instructions and operands, for example a four-byte immediate value can result in around four billion different tokens \cite{ahn2022}. This high number of tokens leads to an out-of-vocabulary (OOV) problem, making it difficult to generate meaningful input embeddings \cite{ahn2022}. DeepSemantic introduced a normalization approach that aimed to preserve the original semantics of the binary code while reducing the number of distinct tokens. However, it was found that after normalization, many initially unique functions were mapped to the same normalised function. As a result, during the de-duplication process, this merging led to a loss of information. 
The Peekaboo DBI tool executed the samples which resulted in thousands of different memory addresses, that were variously used as operands. The different Transformer models vocabulary sizes range from 30,000 to 50,000 which initially resulted in numerous ASM instructions being classified as Out Of Vocabulary (OOV). Consequently, certain memory addresses were replaced with \texttt{memoryaddress}, as shown in Tables \ref{tab:ASM Instruction Transformations}, \ref{tab:Function definitions}. However, immediate values that stand as values themselves, or point to data such as variables, or function as an offset or displacement were not changed. The approaches used by DeepSemantic and BinShot use static disassembly to obtain the ASM instructions and uses a normalization strategy that replaces immediate values that denote a call invocation within or outside the current binary, a jump destination, a library function target, a string reference, or a statically-allocated variable \cite{ahn2022,koo2021}. In contrast, the Peekaboo data was captured during dynamic analysis where the samples were actually executed. Static disassembly lacks context as the binary is not executed and relies soles on the binary’s instructions and metadata, which may not fully resolve all addresses, constants, or data references, particularly if they are computed or resolved dynamically at runtime. Peekaboo captures the ASM instructions during the actual execution and the addresses, constants, and data references are resolved during runtime. The objective is to preserve as much of this valuable information as possible and we found that only replacing specific immediate values resolved the OOV issues. This balance, ensures that the Transformer model can effectively learn and generate embeddings from the normalized representations of binary functions. Further, the ASM instructions and their operands were either treated as individual words or concatenated and treated as a single word as detailed in Table \ref{tab:ASM Instruction Transformations}. 
\begin{table}[t]
\centering
\footnotesize
\captionsetup{justification=centering}
  \caption{ASM Instruction Transformations}
  \label{tab:ASM Instruction Transformations}
  \begin{tabular}{ccc}
    \toprule
     \makecell{Original Instruction} & \makecell{Not Concatenated} & \makecell{Concatenated}\\
    \midrule
        \makecell{mov esp, esi} & \makecell{mov esp, esi} & \makecell{movespesi} \\
        \makecell{pop esi} & \makecell{pop esi} & \makecell{popesi} \\
        \makecell{mov byte ptr [ebp-0x19], al} & \makecell{mov byte ptr [ebp-0x19], al} & \makecell{movbyteptr[ebp-0x19]al} \\
        \makecell{call 0x775ade2d} & \makecell{call memoryaddress} & \makecell{callmemoryaddress} \\
        \makecell{mov eax, dword ptr fs:[0x30]} & \makecell{mov eax, dword ptr fs:[0x30]} & \makecell{moveaxdwordptrfs:[0x30]} \\
    \bottomrule
\end{tabular}
\end{table}

\begin{table}[t]
\centering
\footnotesize
\captionsetup{justification=centering}
  \caption{Function normalisation}
  \label{tab:Function definitions}
  \begin{tabular}{cc}
    \toprule
     \makecell{Peekaboo DBI data} & \makecell{Concatenated and normalised functions}\\
    \midrule
        \makecell{mov esp, esi} & \makecell{}\\
        \makecell{pop ebx} & \makecell{}\\
        \makecell{pop edi} & \makecell{}\\
        \makecell{pop esi} & \makecell{}\\
        \makecell{pop ebp} & \makecell{}\\
        \makecell{\textbf{ret 0x10}} & \makecell{movespesi popebx popedi popesi popebp \textbf{ret0x10}}\\
       \noalign{\vskip 5pt}
        \makecell{mov byte ptr [ebp-0x19], al} & \makecell{}\\
        \makecell{mov dword ptr [ebp-0x4], 0xfffffffe} & \makecell{}\\
        \makecell{mov dword ptr [ebp-0x24], 0x0} & \makecell{movbyteptr[ebp-0x19]al movdwordptr[ebp-0x4]memoryaddress}\\
        \makecell{\textbf{call 0x775ade2d}} & \makecell{movdwordptr[ebp-0x24]0x0 \textbf{callmemoryaddress}}\\
       \noalign{\vskip 5pt}
        \makecell{mov eax, dword ptr fs:[0x30]} & \makecell{}\\
        \makecell{mov eax, dword ptr [eax+0x50]} & \makecell{}\\
        \makecell{test eax, eax} & \makecell{}\\
        \makecell{jnz 0x775a3d2e} & \makecell{moveaxdwordptrfs:[0x30] moveaxdwordptr[eax+0x50]}\\
        \makecell{\textbf{ret}} & \makecell{testeaxeax jnz memoryaddress \textbf{ret}}\\

    \bottomrule
\end{tabular}
\end{table}

There are normally 6 ASM instructions that are used to set up the stack, so after splitting the data into functions those with less than 6 instructions were dropped from the dataset. A function cannot be labelled benign and malicious across different samples. Consequently, the data was de-duplicated and filtered. Any function in a ransomware sample that was also in a benign sample was removed from the ransomware sample. Further, the not concatenated approach split raw Peekaboo data into functions on \texttt{ret} and \texttt{call} and \texttt{push ebp}. The concatenated split the data on \texttt{ret} and \texttt{call}, as shown in Table \ref{tab:Function definitions}.
The various models use different encoding. The three main types of tokenizers used in Transformers are Byte-Pair Encoding (BPE), WordPiece, and SentencePiece. Standard tokenizers were used for experiment A and tokenizers trained on our binary corpus were used in experiment B. 

\subsection{Tokenizer Training}
The bert-base-uncased and distilbert-base-uncased tokenizers use WordPiece tokenization and sub word tokens to handle out of vocabulary words, this allows BERT to handle unseen words by representing them as combinations of more common subwords \cite{devlin2019}. The \#\# symbols indicate that the preceding characters are part of a larger subword token. For example, \#\#v following mo indicates that v is part of the same word token, that is mov, as shown in Table \ref{tab:tokenizer details}. 

DistilRoberta and GPT-2 use Byte Pair Encoding (BPE) \cite{liu2019}. This technique breaks down words into subword units based on the frequency of their occurrence in the training data \cite{liu2019}. The presence of Ġ indicates a space or control character in the tokenized sequence, which helps the model understand where word boundaries occur. The Ġ symbol is similar to how BERT's tokenizer uses \#\# to denote subword boundaries. BPE tokenization, which merges frequent sequences of characters into a single token, allows the model to handle a larger vocabulary and improves performance on tasks requiring nuanced understanding of text. Like the BERT tokenizer, spaces and symbols, for example [ ] , are treated as individual tokens, preserving their integrity within the tokenized sequence, as shown in Table \ref{tab:tokenizer details}. The standard tokenizers used subword tokens extensively indicating a large number of the ASM instructions were OOV. However, the custom tokenizers that were trained on our normalized functions did not suffer from OOV issues.


\begin{table}[t]
\centering
\footnotesize
\captionsetup{justification=centering}
  \caption{Tokenizer details}
  \label{tab:tokenizer details}
  \begin{tabular}{cccc}
    \toprule
     \makecell{Tokenizer} & \makecell{C} & \makecell{Vocab} & \makecell{Tokens}\\
    \midrule
 
        \makecell{WordPiece\\bert-base\\-uncased} & \makecell{N} & \makecell{30,000} & \makecell{mo \#\#v ea \#\#x , d \#\#word\\ pt \#\#r f \#\#s : [ 0 \#\#x \#\#30 ]\\mo \#\#v ea \#\#x , d \#\#word\\ pt \#\#r [ ea \#\#x + 0 \#\#x \#\#50 ]\\test ea \#\#x , ea \#\#x j \#\#nz\\ memory \#\#ad \#\#dre \#\#ss re \#\#t\\}\\
\noalign{\vskip 5pt}

         \makecell{WordPiece\\distilbert-base\\-uncased} & \makecell{N} & \makecell{30,000} & \makecell{mo \#\#v ea \#\#x , d \#\#word\\ pt \#\#r f \#\#s : [ 0 \#\#x \#\#30 ]\\mo \#\#v ea \#\#x , d \#\#word\\ pt \#\#r [ ea \#\#x + 0 \#\#x \#\#50 ]\\test ea \#\#x , ea \#\#x j \#\#nz\\ memory \#\#ad \#\#dre \#\#ss re \#\#t} \\
\noalign{\vskip 5pt}
        \makecell{SentencePieceBPE\\distilroberta\\-base} & \makecell{N} & \makecell{50,000} & \makecell{m ov Ġe ax , Ġd word Ġptr Ġfs :[ 0 x 30 ]\\Ġmov Ġe ax , Ġd word Ġptr Ġ[ e ax + 0 x 50 ]\\Ġtest Ġe ax , Ġe ax Ġj nz Ġmemory address Ġret} \\
\noalign{\vskip 5pt}
        \makecell{GPT-2-BPE} & \makecell{N} & \makecell{50,257} & \makecell{move ax d word ptr fs :[ 0 x 30 ]\\Ġmove ax d word ptr [ e ax + 0 x 50 ]\\Ġtest e axe ax Ġj nz Ġmemory address Ġret} \\

\noalign{\vskip 5pt}
        \makecell{WordPiece\\distilbert-custom} & \makecell{Y} & \makecell{30,522} & \makecell{moveaxdwordptrfs : [ 0x30 ]\\moveaxdwordptr [ eax + 0x50 ]\\testeaxeax jn \#\#z memoryaddress ret} \\

\noalign{\vskip 5pt}
        \makecell{SentencePieceBPE\\distilroberta-custom} & \makecell{Y} & \makecell{50,000} & \makecell{\_moveaxdwordptrfs:[0x30]\\\_moveaxdwordptr[eax+0x50]\\\_testeaxeax \_jnz \_ memoryaddress \_ret} \\
\noalign{\vskip 5pt}
        \makecell{GPT2-custom} & \makecell{Y} & \makecell{50,257} & \makecell{moveaxdwordptrfs :[ 0 x 30 ]\\ Ġmoveaxdwordptr [ eax + 0 x 50 ]\\Ġtesteaxeax Ġjnz Ġ memoryaddress Ġret} \\
\noalign{\vskip 5pt}
        \makecell{SentencePiece\\XLNet-custom} & \makecell{Y} & \makecell{50,257} & \makecell{\_move ax d word pt r f s : [ 0 x 30 ]\\ \_move ax d word pt r [ e ax + 0 x 50 ]\\ \_test e ax e ax \_ j nz \_memory address \_ re t} \\
        
    \bottomrule
\end{tabular}
\end{table}



\begin{table}[ht]
\centering
\footnotesize
\caption{Training testing dataset details}
\label{tab:Training testing dataset details}
\begin{tabular}{cccccc}
\hline
\makecell{Experiment} & \multicolumn{2}{c}{\makecell{Training}} & \multicolumn{2}{c}{\makecell{Testing}} \\
\hline
\multirow{13}{*}{\makecell{A}} & BlackCat & 62 & LockBit & 26 \\
& Chaos & 2 & Maze & 9 \\
& Clop & 2 & Petya & 5 \\
& Conti & 51 & WannaCry & 48 \\
& DarkSide & 47 & Benign & 21 \\
& Dharma & 10 & & \\
& Hive & 15 & & \\
& Locky & 17 & & \\
& NetWalker & 33 & & \\
& RagnarLocker & 2 & & \\
& Ryuk & 26 & & \\
& Sodinokibi & 70 & & \\
& Stop & 80 & & \\
& Benign & 287 & & \\
\\
\multirow{2}{*}{\makecell{B}} & \makecell{as per A} & \makecell{as per A} & \makecell{as per A} & \makecell{as per A} \\
& Benign & 277 & Benign & 25 \\
\hline
\end{tabular}
\end{table}

\begin{table}[h]
\centering
\footnotesize
\captionsetup{justification=centering}
  \caption{Transformer models hyperparameters}
  \label{tab:Hyperparameters}
  \begin{tabular}{ccccc}
    \toprule
     \makecell{Model} & \makecell{Layers} & \makecell{Hidden Layers} & \makecell{Attention Heads} & \makecell{Parameters}\\
    \midrule
        \makecell{BERTBase} & \makecell{12 Encoder} & \makecell{768} & \makecell{12} & \makecell{110M}\\
        \makecell{DistilBERT} & \makecell{6 Encoder} & \makecell{768} & \makecell{12} & \makecell{66M}\\
        \makecell{DistilRoBERTa} & \makecell{6 Encoder} & \makecell{768} & \makecell{12} & \makecell{82M}\\
        \makecell{GPT2 small} & \makecell{12 Decoder} & \makecell{768} & \makecell{12} & \makecell{1.5B}\\
        \makecell{XLNet small} & \makecell{12 Encoder} & \makecell{768} & \makecell{12} & \makecell{110M}\\        
    \bottomrule
\end{tabular}
\end{table}

\section{Experimental Analysis}
\label{sec:results}
To determine the effectiveness of Transformer models with the Peekaboo DBI data two experiments were performed. Experiment A used various Transformer models, the standard tokenizers, and the normalized functions that were not concatenated. Experiment B used various Transformer models, custom tokenizers, and concatenated normalized functions. The train test samples are detailed in Table \ref{tab:Training testing dataset details} and the hyperparameters for the various transformer models are shown in Table \ref{tab:Hyperparameters}. For more detail please see the source code at \cite{gabertpro2024}.
\subsection{Experiment setup and evaluation}
In this section, we detail the computational environment and performance metrics used. Google Colaboratory \cite{googlecolab} was utilized to train and test the Transformer models, leveraging Python 3 to execute the experiments. Various GPUs, depending on availability, were used for processing and optimization of the models. This setup facilitated comprehensive evaluation and fine-tuning of the models in a cloud-based environment.

To assess the accuracy of the Transformer models and their ability to identify and classify both ransomware and benign software correctly, we compute precision, recall, and F1 scores. Accuracy measures the overall proportion of correct predictions, but can be misleading in cases of class imbalance. Precision indicates the proportion of true ransomware predictions out of all predictions made, highlighting the model's correctness in predicting ransomware samples. Recall (or Sensitivity) measures the proportion of actual ransomware samples correctly identified by the model, focusing on how well it detects ransomware samples. The F1 score combines precision and recall into a single metric, providing a balanced measure of the model's performance when precision and recall are in a trade-off.
\subsection{Experiment A}
Table \ref{tab:Experiment A training Corpus} shows the detailed breakdown of the number of functions, for the Peekaboo ransomware and benign samples combined, as detailed in Table \ref{tab:Training testing dataset details}. As a comparison the BinShot and DeepSemantic data is also presented \cite{ahn2022,koo2021}. However, the binaries were very different, the Peekaboo data is live ransomware and benign software whereas BinShot and DeepSemantic were focused on BCSD using GNU utilities, SPEC2006 and SPEC2017 with different compilers and optimizations \cite{ahn2022,koo2021}. Further, Peekaboo used DBI whereas BinShot and DeepSemantic used static disassembly \cite{gaber2024, ahn2022,koo2021}.  As described in Section~\ref{sec:methods} duplicate functions were removed and any function present in the malware samples that also appeared in the benign samples was removed from the malicious dataset. Of these filtered functions 55.25\% were longer than 256 tokens, this was because the ASM instructions and operands were treated as individual words in this experiment.
\begin{table}[t]
\centering
\footnotesize
\captionsetup{justification=centering}
  \caption{Experiment A Normalised functions training corpus}
  \label{tab:Experiment A training Corpus}
  \begin{tabular}{cccccc}
    \toprule
     \makecell{Dataset} & \makecell{Binaries} & \makecell{Initial} & \makecell{Deduplicated} & \makecell{Filtered} & \makecell{\%\textgreater256}\\
    \midrule
        \makecell{Peekaboo} & \makecell{704} & \makecell{13,201,202} & \makecell{692,980} & \makecell{589,176}& \makecell{55.25}\\
        
        \makecell{BinShot} & \makecell{1,400} & \makecell{1,770,675} & \makecell{18,499} & \makecell{18,499}& \makecell{16.74}\\
        \makecell{DeepSemantic} & \makecell{1,328} & \makecell{1,690,715} & \makecell{17,220} & \makecell{17,220}& \makecell{na}\\
        
    \bottomrule
\end{tabular}
\end{table}

\begin{figure}[H]
    \centering
    \begin{minipage}{0.43\textwidth}
        \centering
        \includegraphics[width=\textwidth]{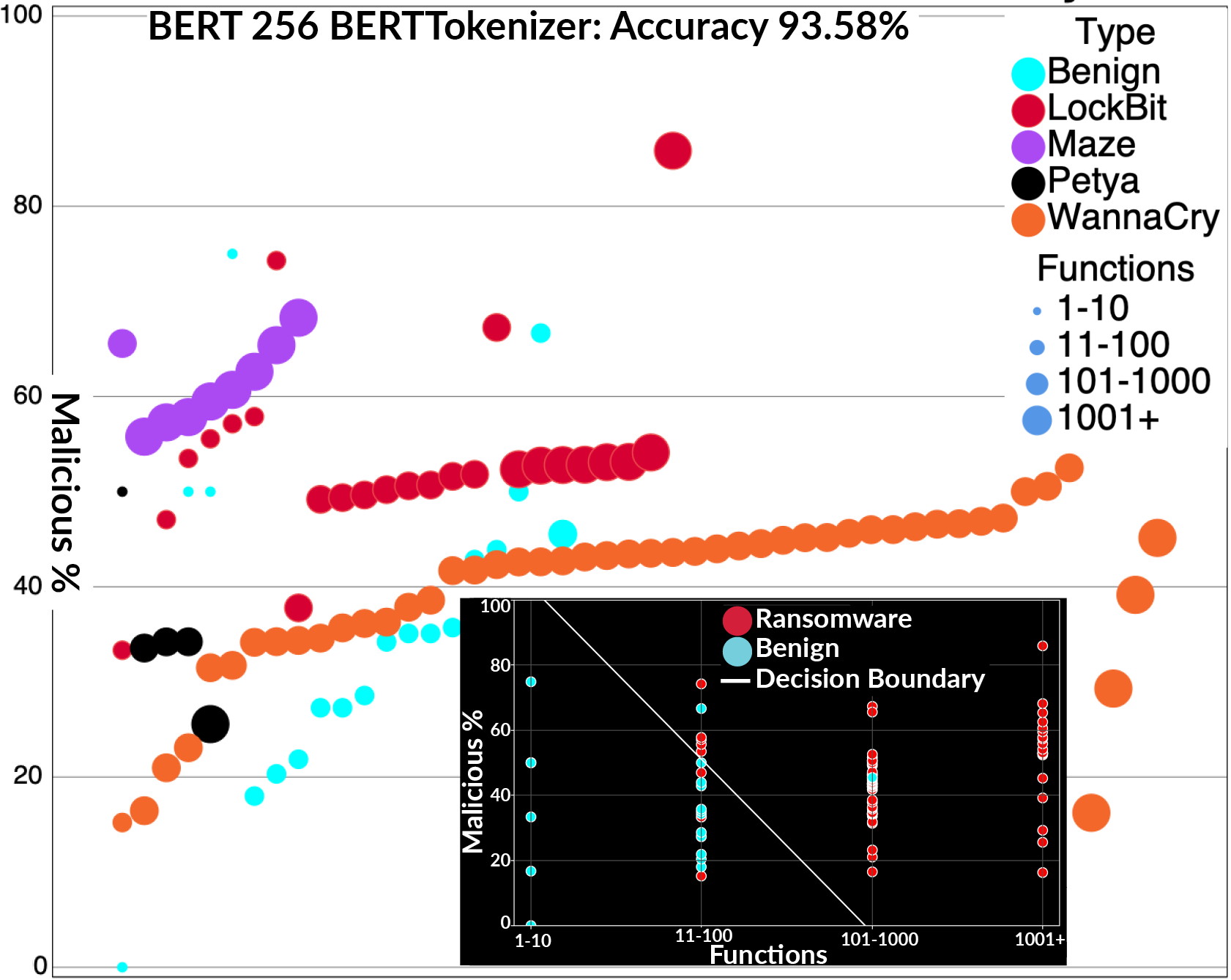}
    \end{minipage}
    \begin{minipage}{0.43\textwidth}
        \centering
        \includegraphics[width=\textwidth]{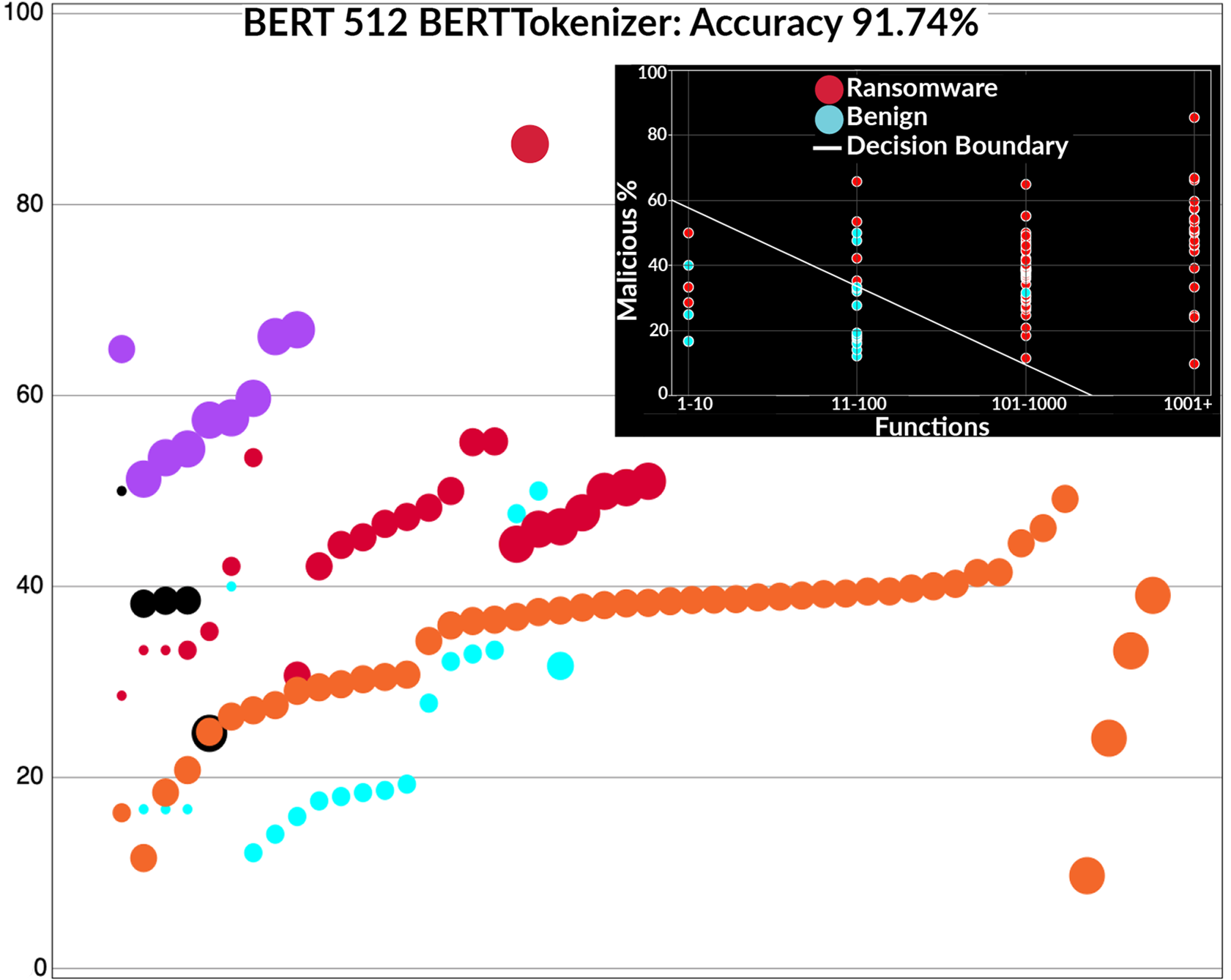}
    \end{minipage}


    \begin{minipage}{0.43\textwidth}
        \centering
        \includegraphics[width=\textwidth]{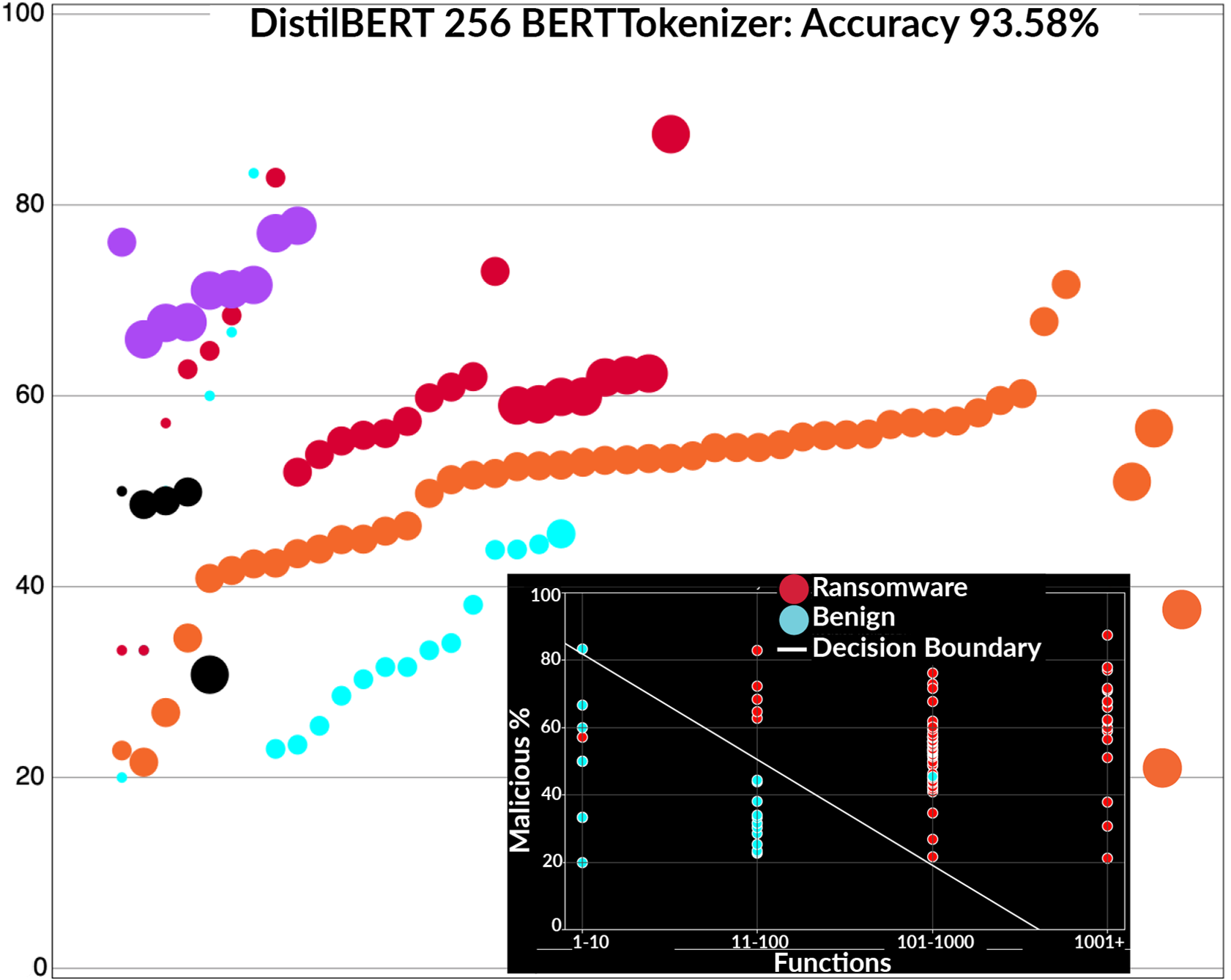}
    \end{minipage}
    \begin{minipage}{0.43\textwidth}
        \centering
        \includegraphics[width=\textwidth]{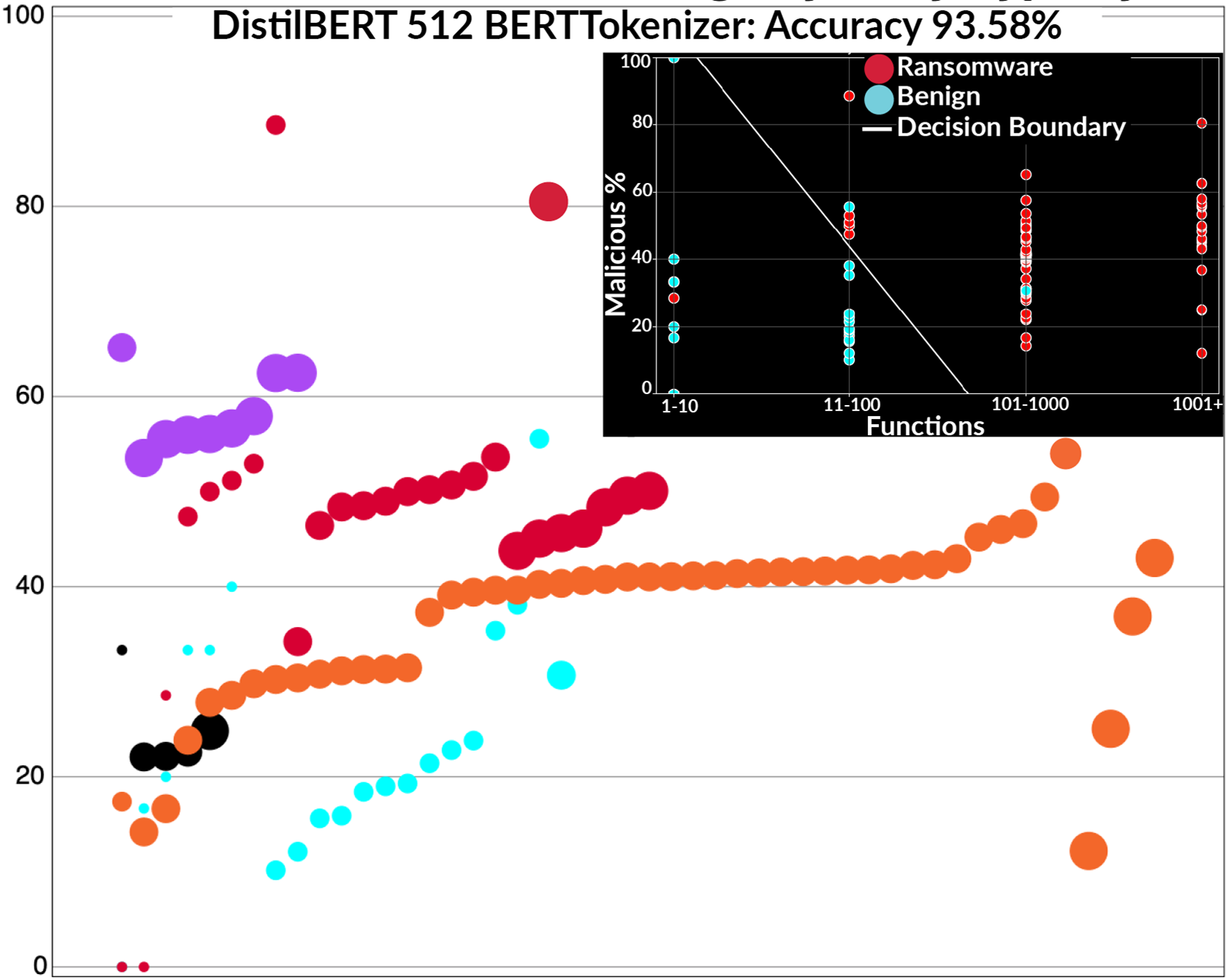}
    
    \end{minipage}


    \begin{minipage}{0.43\textwidth}
        \centering
        \includegraphics[width=\textwidth]{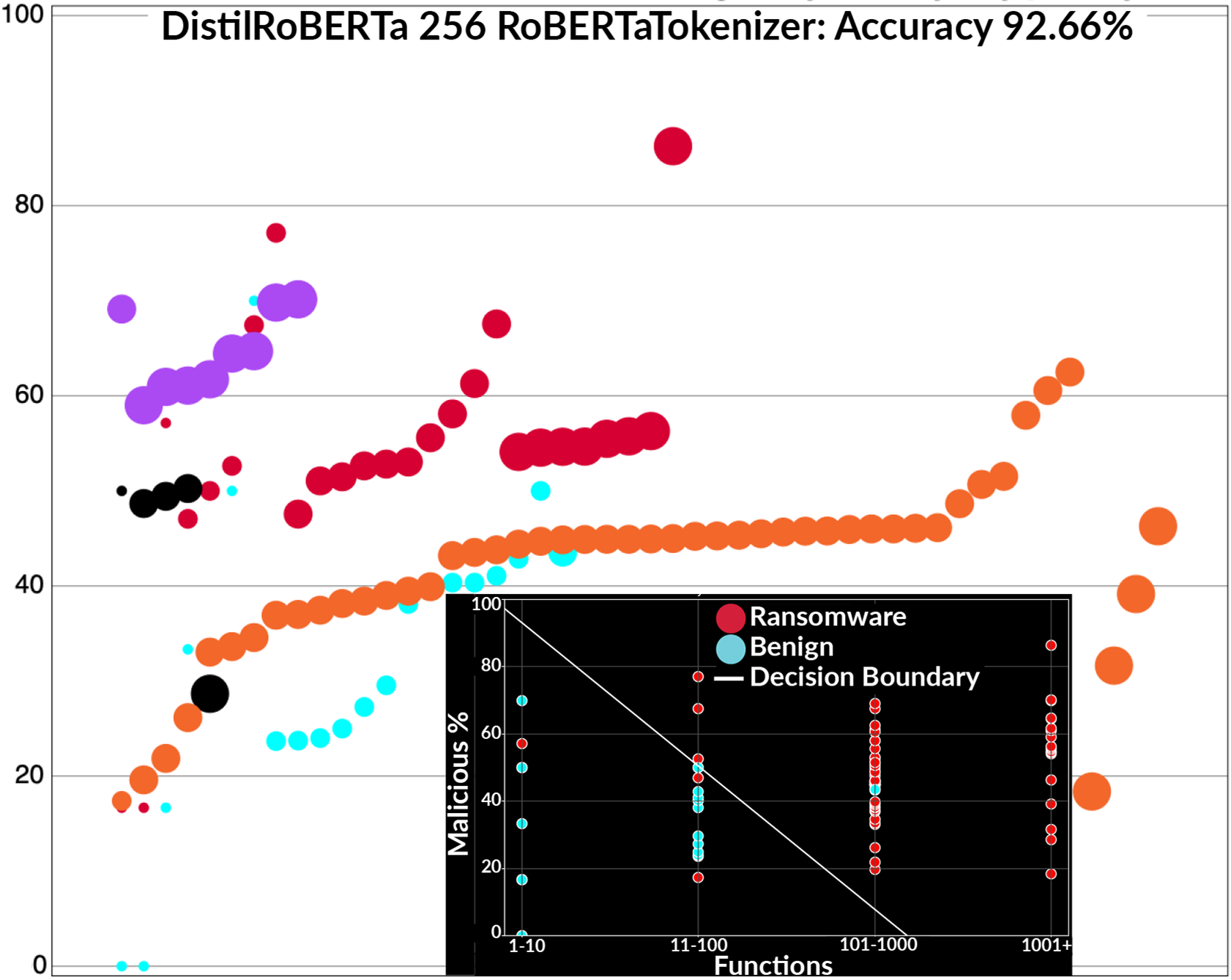}
    
    \end{minipage}
    \begin{minipage}{0.43\textwidth}
        \centering
        \includegraphics[width=\textwidth]{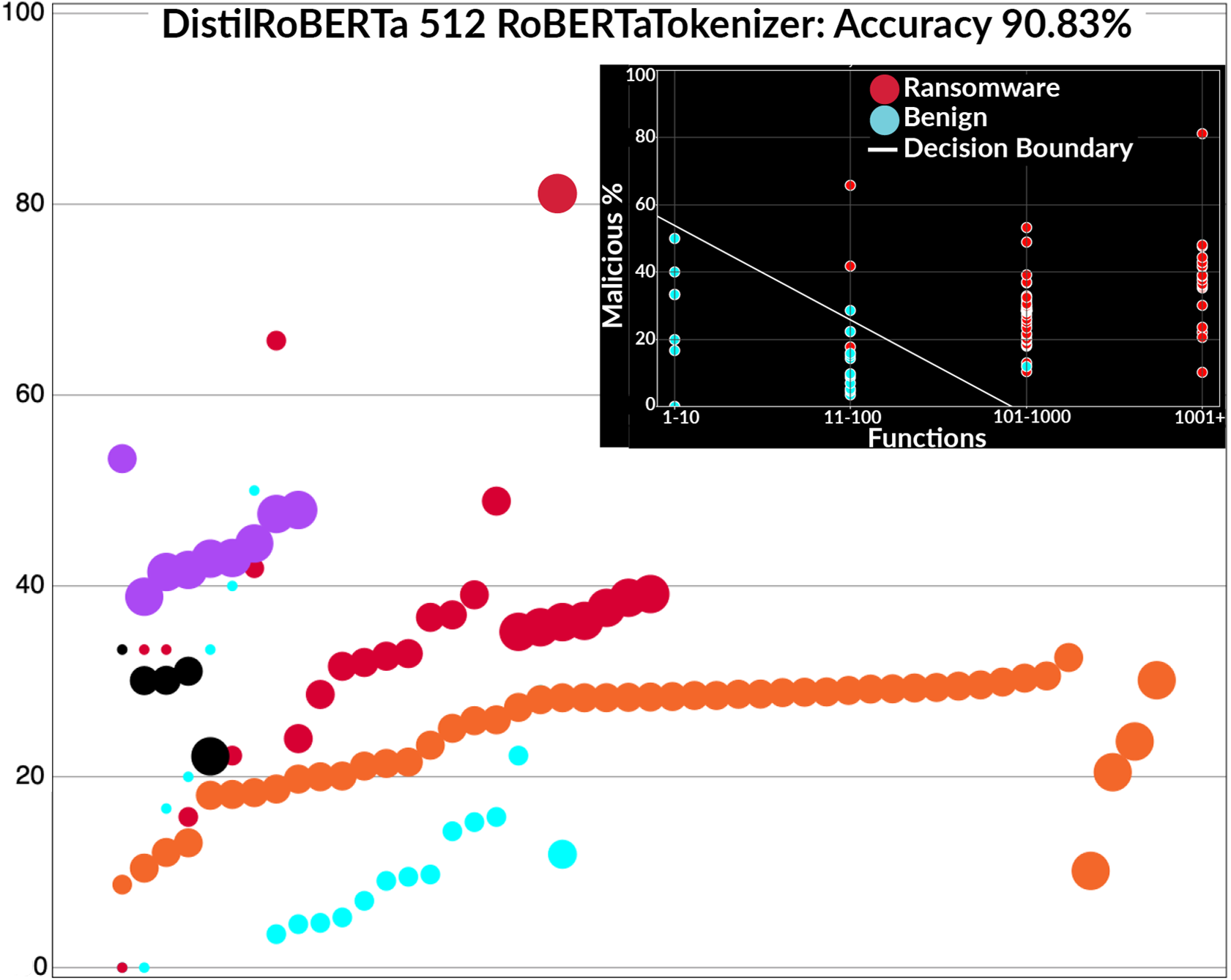}
        
    \end{minipage}

    \begin{minipage}{0.43\textwidth}
        \centering
        \includegraphics[width=\textwidth]{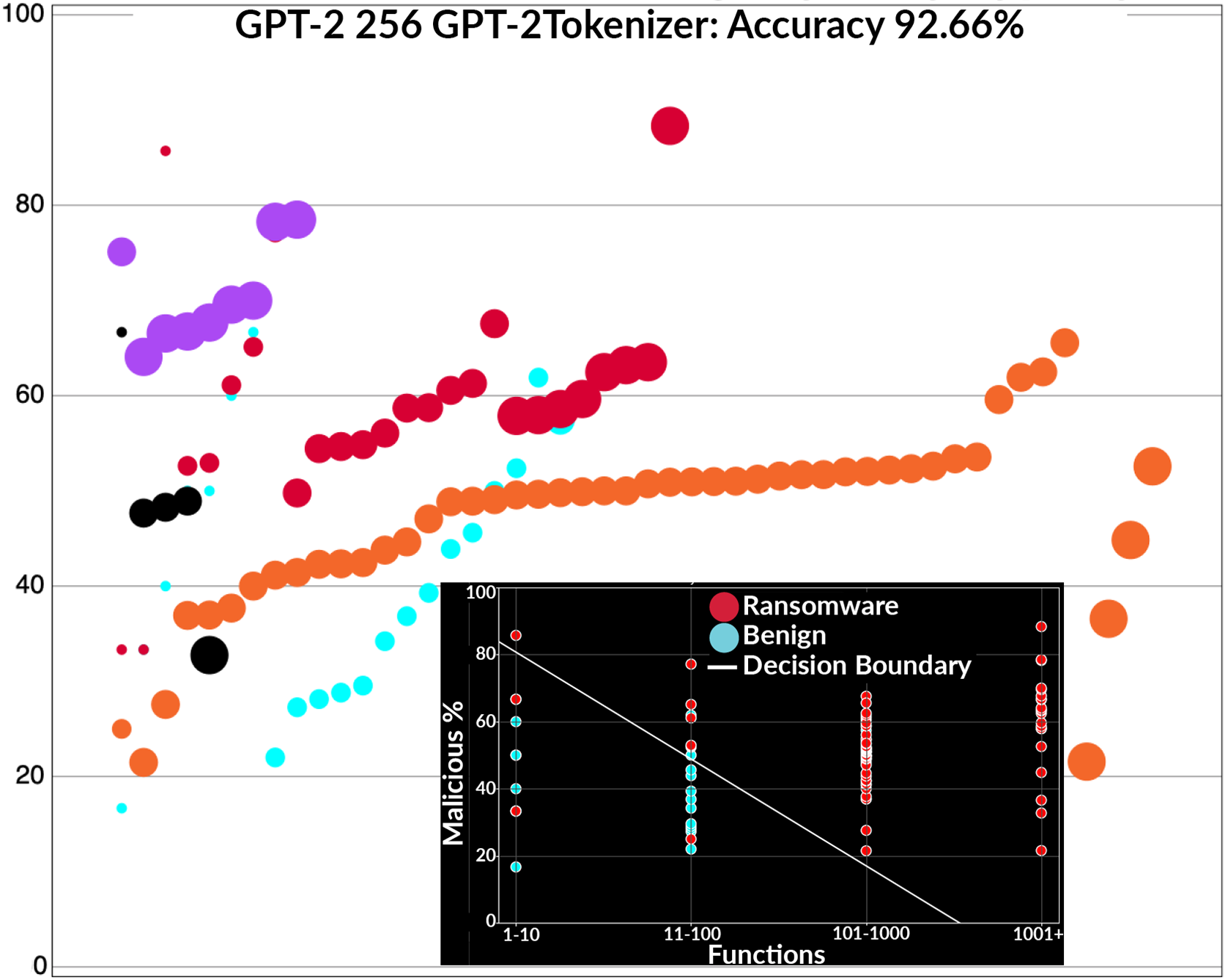}
        
    \end{minipage}
    \caption{Experiment A results}
    \label{fig:expagroupedimages}
\end{figure}

The fine tuning parameters and validation results for the models are shown in Table \ref{tab:finetuningA}. A batch size of 8 was used with GPT-2 to achieve a reasonable training time. As detailed in Section \ref{sec:methods} the classification layer predicted the number of benign functions and malicious functions for each sample. The output layer performed the final classification for the test samples by utilizing a Support Vector Machine (SVM) hyperplane, which acts as the decision boundary to delineate between benign and malicious samples, as detailed in Figure \ref{fig:System design of Peekaboo Classifier}. That is, the models classify the individual functions and a SVM hyperplane is used to classify the samples, as shown in Figure \ref{fig:expagroupedimages}. The test samples detailed in Table \ref{tab:Training testing dataset details} were used to evaluate the various Transformer models in terms of their ability to classify never before seen malicious behaviours. Recall that any function present in the test samples that also appeared in the training data was removed from the test samples. The classification layer predicted the number of benign functions and malicious functions for each sample which is presented as malicious \% in Figure \ref{fig:expagroupedimages}. The size of the bubble reflects how many functions the samples contained and ranged from 3 to more than a 1000, separated into 4 classes. The insets show the SVM decision boundary used to separate the benign and ransomware samples and the performance metrics are presented in Table \ref{tab:expamodel_performance}. All models achieved over 90\% accuracy however the decision boundary intercepts the x axis within the data range which results in misleading accuracies, this is because there was only one benign sample with 100-1000 functions and none with more than 1000+ functions. For Experiment B the dataset was re balanced to include benign samples with 100-1000 and 1000+ instructions in the test set. 

\begin{table}[t]
\centering
\footnotesize
\captionsetup{justification=centering}
  \caption{Experiment A Model and fine tuning details\\ \textbf{Note:} nc=not concatenated}
  \label{tab:finetuningA}
  \begin{tabular}{ccccccc}
    \toprule
    \makecell{Model} & \makecell{Epochs} & \makecell{Batch Size} & \makecell{Tokenizer} & \makecell{Word}& \makecell{Validation\\Loss}& \makecell{Validation\\Accuracy \%}\\
    \midrule
        \makecell{BERT 256} & \makecell{3} & \makecell{16} & \makecell{bert-base-uncased} & \makecell{nc}& \makecell{0.3063}& \makecell{84.44}\\
        \makecell{BERT 512} & \makecell{3} & \makecell{16} & \makecell{bert-base-uncased} & \makecell{nc} & \makecell{0.3015} & \makecell{85.10} \\
        \makecell{DistilBERT 256} & \makecell{3} & \makecell{16} & \makecell{distilbert-base-uncased} & \makecell{nc} & \makecell{0.2984} & \makecell{84.81} \\
        \makecell{DistilBERT 512} & \makecell{3} & \makecell{16} & \makecell{distilbert-base-uncased} & \makecell{nc} & \makecell{0.2911} & \makecell{84.94} \\
        \makecell{DistilRoBERTa 256} & \makecell{3} & \makecell{16} & \makecell{distilroberta-base} & \makecell{nc} & \makecell{0.3012} & \makecell{83.02} \\
        \makecell{DistilRoBERTa 512} & \makecell{3} & \makecell{16} & \makecell{distilroberta-base} & \makecell{nc} & \makecell{0.2994} & \makecell{84.13} \\
        \makecell{GPT2 256} & \makecell{3} & \makecell{8} & \makecell{gpt2} & \makecell{nc} & \makecell{0.038} & \makecell{84.10} \\
    \bottomrule
\end{tabular}
\end{table}

\begin{table}[t]
\centering
\footnotesize
\captionsetup{justification=centering}
\caption{Experiment A performance metrics}
\label{tab:expamodel_performance}
\begin{tabular}{lcccc}
    \toprule
    \makecell{Model} & \makecell{Accuracy} & \makecell{Precision} & \makecell{Recall} & \makecell{F1 Score} \\
    \midrule
    \makecell{BERT 256 BERTTokenizer} & \makecell{93.58} & \makecell{97.65} & \makecell{94.32} & \makecell{95.96} \\
    \makecell{BERT 512 BERTTokenizer} & \makecell{91.74} & \makecell{96.47} & \makecell{93.18} & \makecell{94.81} \\
    \makecell{DistilBERT 256 BERTTokenizer} & \makecell{93.58} & \makecell{97.65} & \makecell{94.32} & \makecell{95.96} \\
    \makecell{DistilBERT 512 BERTTokenizer} & \makecell{93.58} & \makecell{97.65} & \makecell{94.32} & \makecell{95.96} \\
    \makecell{DistilRoBERTa 256 RobertaTokenizer} & \makecell{92.66} & \makecell{98.78} & \makecell{92.05} & \makecell{95.27} \\
    \makecell{DistilRoBERTa 512 RobertaTokenizer} & \makecell{90.83} & \makecell{97.56} & \makecell{90.91} & \makecell{94.08} \\
    \makecell{GPT-2 256 GPT-2Tokenizer} & \makecell{92.66} & \makecell{95.45} & \makecell{95.45} & \makecell{95.45} \\
    \bottomrule
\end{tabular}
\end{table}

\subsection{Experiment B}
\begin{figure}[h!]
    \centering
    \begin{minipage}[b]{0.45\textwidth}
        \centering
        \includegraphics[width=\textwidth]{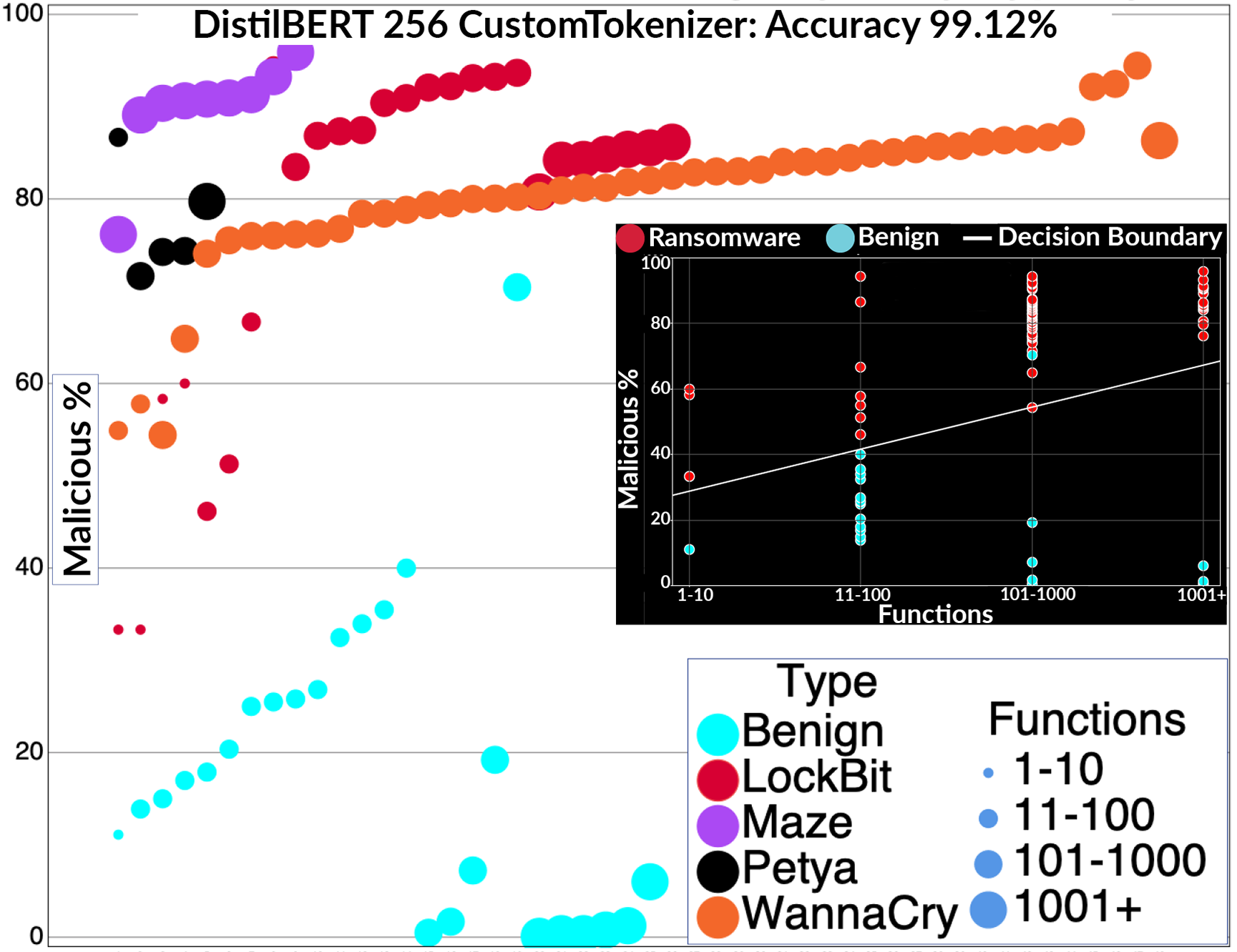}
    
    \end{minipage}
    \hspace{0.01\textwidth}
    \begin{minipage}[b]{0.45\textwidth}
        \centering
        \includegraphics[width=\textwidth]{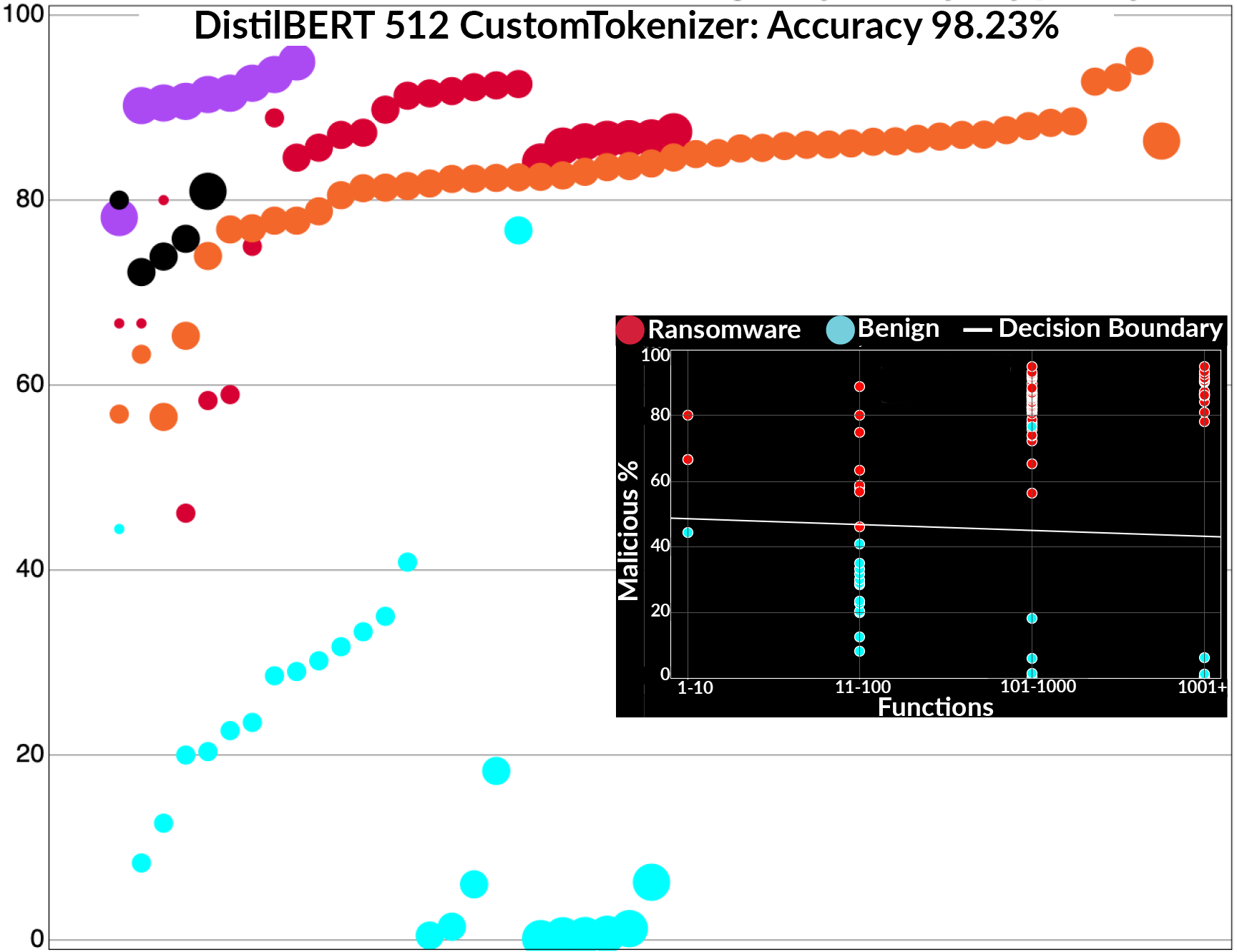}
        
    \end{minipage}

    \vspace{0.01\textwidth} 

    \begin{minipage}[b]{0.45\textwidth}
        \centering
        \includegraphics[width=\textwidth]{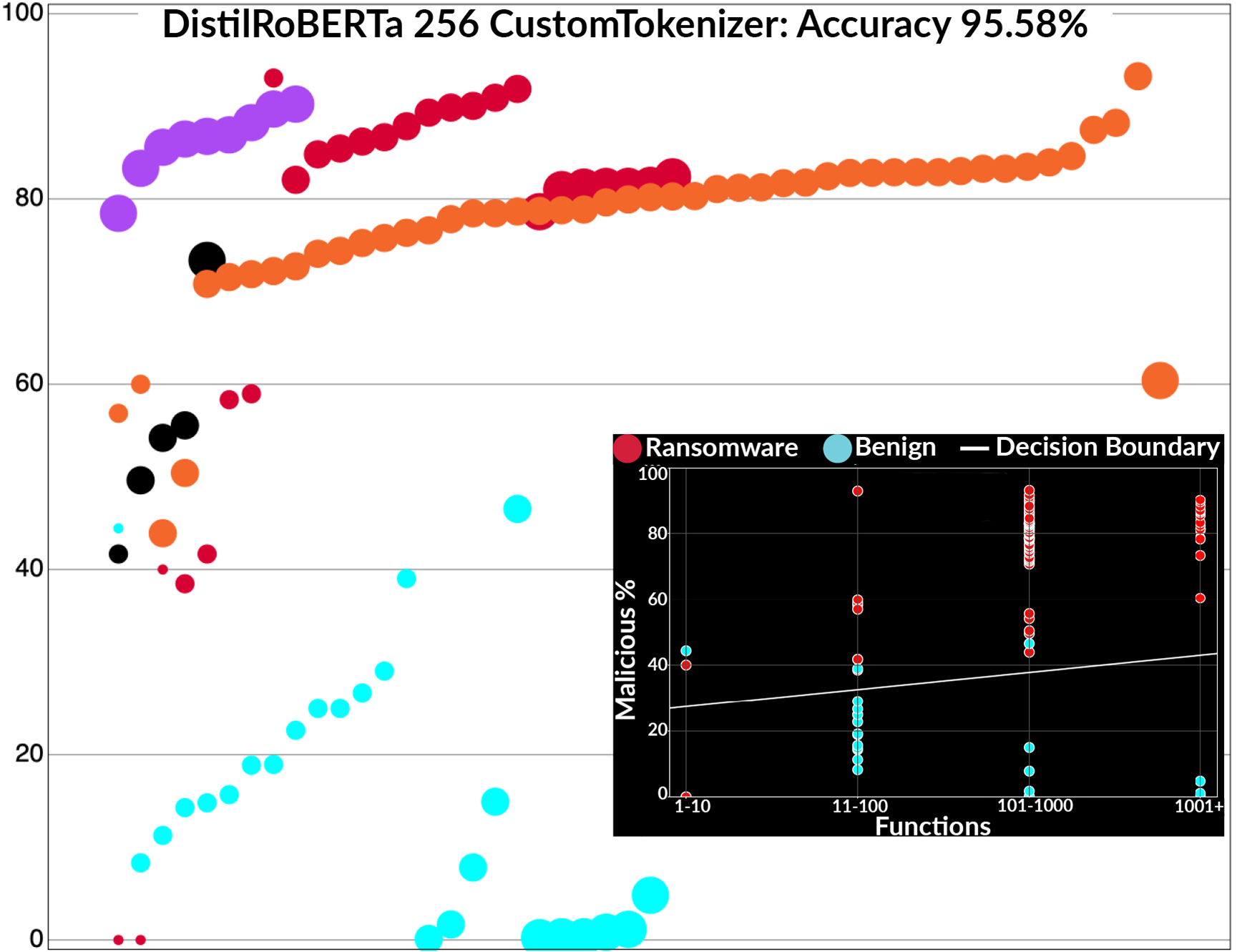}
        
    \end{minipage}
    \hspace{0.01\textwidth}
    \begin{minipage}[b]{0.45\textwidth}
        \centering
        \includegraphics[width=\textwidth]{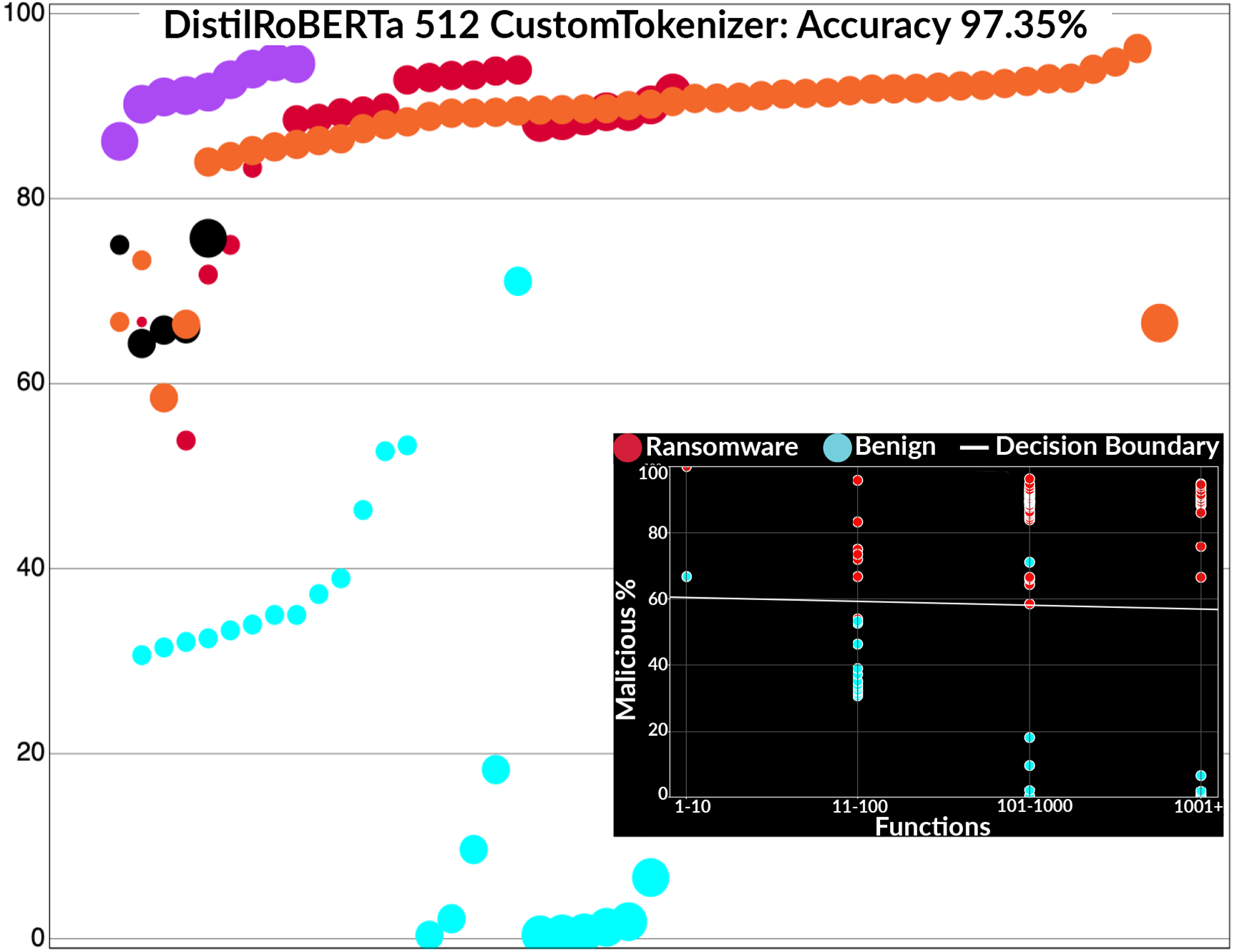}
        
    \end{minipage}

    \vspace{0.01\textwidth} 

    \begin{minipage}[b]{0.45\textwidth}
        \centering
        \includegraphics[width=\textwidth]{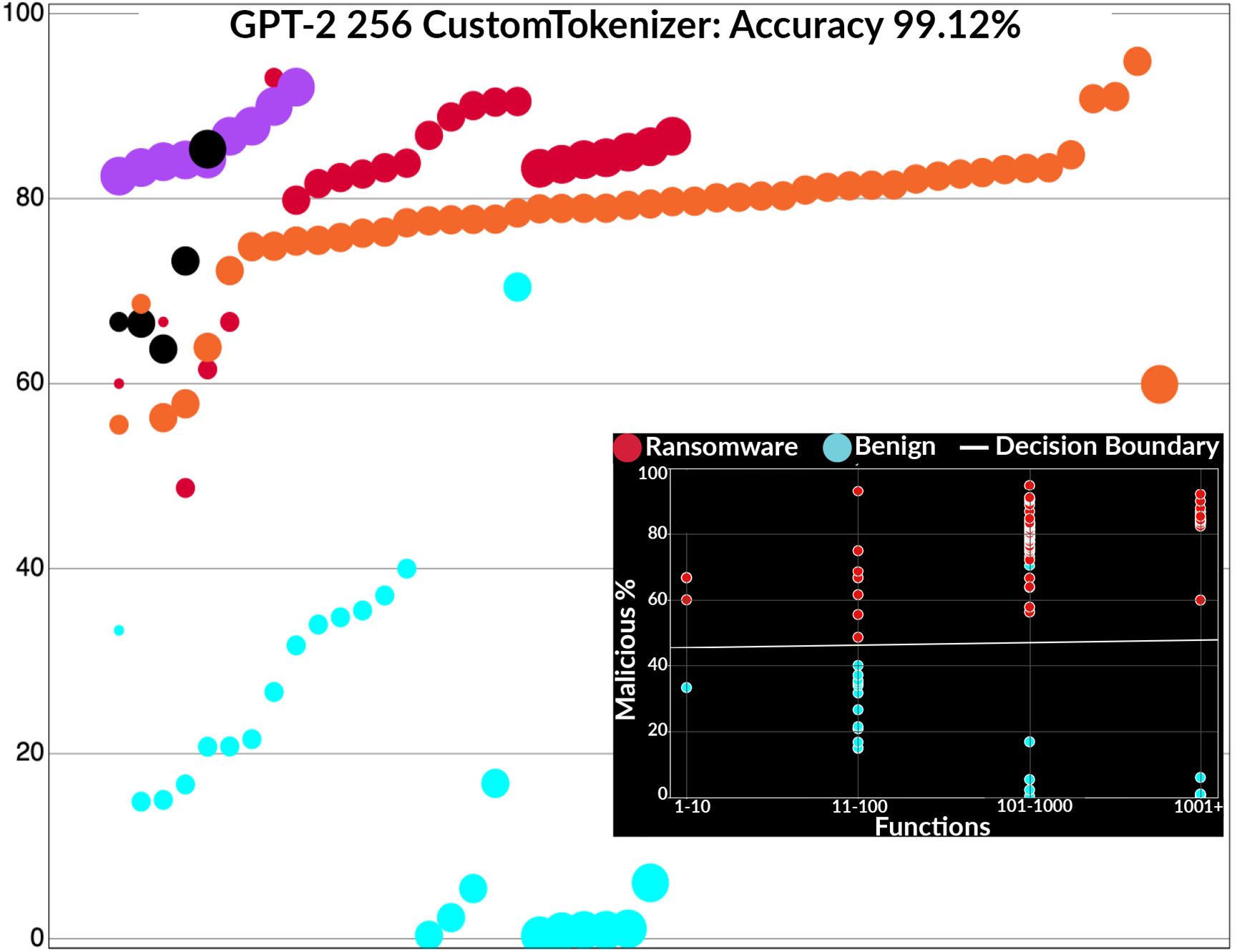}
        
    \end{minipage}
    \hspace{0.01\textwidth}
    \begin{minipage}[b]{0.45\textwidth}
        \centering
        \includegraphics[width=\textwidth]{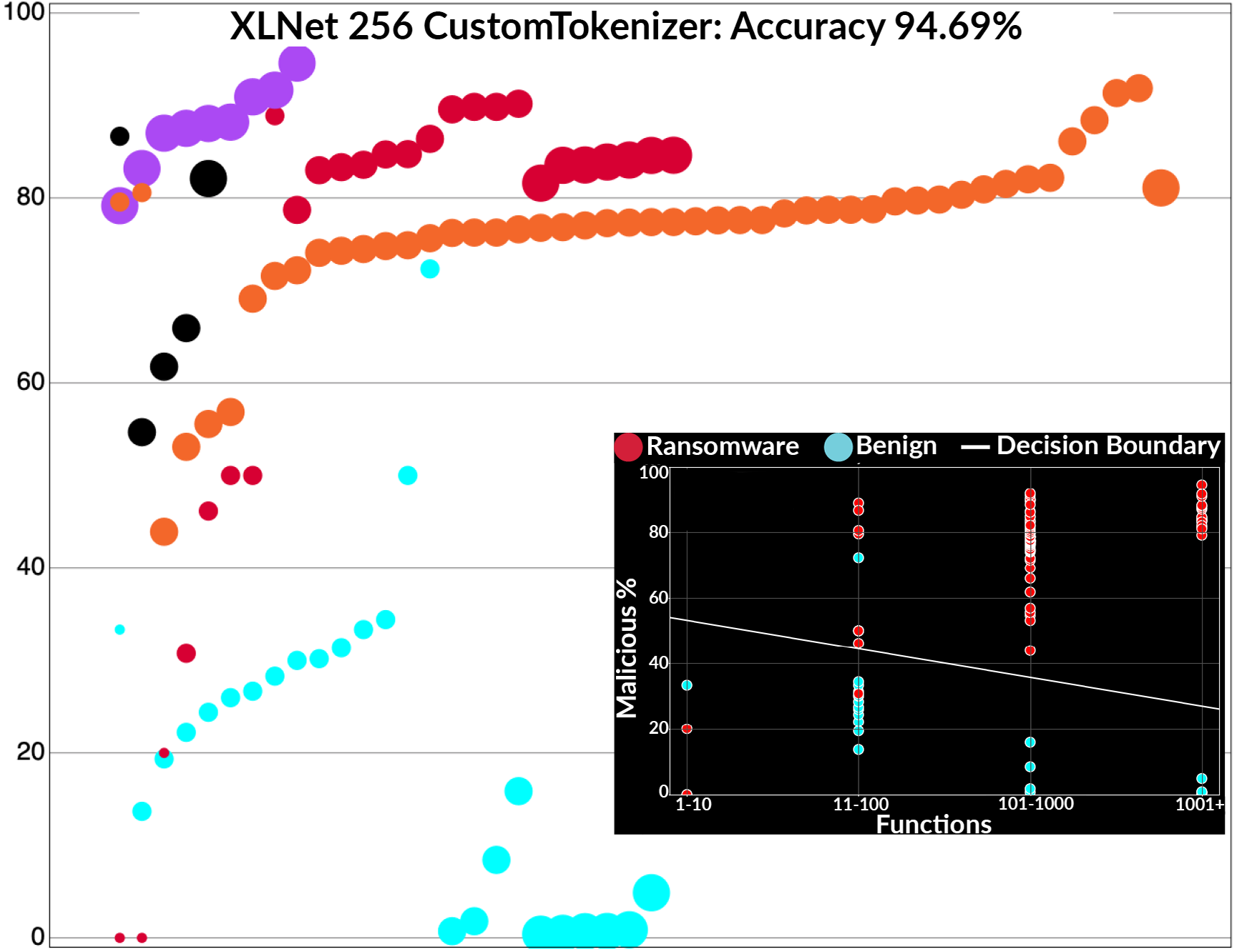}
        
    \end{minipage}

    \vspace{0.01\textwidth} 
    \caption{Experiment B results}
    \label{fig:expbgroupedimages}
\end{figure}

Table \ref{tab:Experiment B training Corpus} shows the detailed breakdown of the number of functions, for the Peekaboo ransomware and benign samples, as detailed in Table \ref{tab:Training testing dataset details}. It is also worth noting the number of processed tokens that were longer than 256 was 2.54\%, which was less than Experiment A. This is because the ASM instructions and operands were concatenated and treated as one word in this experiment. As described in Section~\ref{sec:methods} duplicate functions were removed and any function present in the malware samples that also appeared in the benign samples was removed from the malicious dataset. 
\begin{table}[h!]
\centering
\footnotesize
\captionsetup{justification=centering}
  \caption{Experiment B training corpus}
  \label{tab:Experiment B training Corpus}
  \begin{tabular}{cccccc}
    \toprule
     \makecell{Dataset} & \makecell{Binaries} & \makecell{Initial} & \makecell{Deduplicated} & \makecell{Filtered} & \makecell{\%\textgreater256}\\
    \midrule
        \makecell{Peekaboo} & \makecell{704} & \makecell{5,789,488} & \makecell{786,766} & \makecell{757,141}& \makecell{2.54}\\  
    \bottomrule
\end{tabular}
\end{table}
In this experiment the dataset was re-balanced to include benign samples with 100-1000 and 1000+ instructions in the test set, as detailed in Table \ref{tab:Training testing dataset details}. Further, custom tokenizers were trained on the normalized functions that used ASM instructions and operands concatenated into a single word. The fine tuning parameters and validation results are shown in Table \ref{tab:finetuningB}, achieving a significant performance improvement over experiment A. The classification layer and output layer functionality were unchanged. The results are presented in Figure \ref{fig:expbgroupedimages} and the performance metrics in Table \ref{tab:expbmodel_performance}. All models achieved over 95\% accuracy and the decision boundary is balanced. 

\begin{table}[h!]
\centering
\footnotesize
\captionsetup{justification=centering}
  \caption{Experiment B fine tuning details\\ \textbf{Note:} c=concatenated}
  \label{tab:finetuningB}
  \begin{tabular}{ccccccc}
    \toprule
    \makecell{Model} & \makecell{Epochs} & \makecell{Batch Size} & \makecell{Tokenizer} & \makecell{Word}& \makecell{Validation\\Loss}& \makecell{Validation\\Accuracy \%}\\
    \midrule
    \makecell{DistilBERT 256} & \makecell{3} & \makecell{16} & \makecell{custom} & \makecell{c} & \makecell{0.1008} & \makecell{95.37} \\
    \makecell{DistilBERT 512} & \makecell{3} & \makecell{16} & \makecell{custom} & \makecell{c} & \makecell{0.0962} & \makecell{95.59} \\
    \makecell{DistilRoBERTa 256} & \makecell{3} & \makecell{16} & \makecell{custom} & \makecell{c} & \makecell{0.1103} & \makecell{94.77} \\
    \makecell{DistilRoBERTa 512} & \makecell{3} & \makecell{16} & \makecell{custom} & \makecell{c} & \makecell{0.1073} & \makecell{95.02} \\
    \makecell{GPT-2 256} & \makecell{3} & \makecell{8} & \makecell{custom} & \makecell{c} & \makecell{0.013} & \makecell{95.3} \\
    \makecell{XLNet 256} & \makecell{3} & \makecell{16} & \makecell{custom} & \makecell{c} & \makecell{0.0955} & \makecell{95.35} \\

    \bottomrule
\end{tabular}
\end{table}

\begin{table}[t]
\centering
\footnotesize
\captionsetup{justification=centering}
\caption{Experiment B performance metrics}
\label{tab:expbmodel_performance}
\begin{tabular}{lcccc}
    \toprule
    \makecell{Model} & \makecell{Accuracy} & \makecell{Precision} & \makecell{Recall} & \makecell{F1 Score} \\
    \midrule
    \makecell{DistilBERT 256 CustomTokenizer} & \makecell{99.12} & \makecell{98.88} & \makecell{100} & \makecell{99.44} \\
    \makecell{DistilBERT 512 CustomTokenizer} & \makecell{98.23} & \makecell{98.86} & \makecell{98.86} & \makecell{98.86} \\
    \makecell{DistilRoBERTa 256 CustomTokenizer} & \makecell{95.58} & \makecell{96.63} & \makecell{97.73} & \makecell{97.17} \\
    \makecell{DistilRoBERTa 512 CustomTokenizer} & \makecell{97.35} & \makecell{97.75} & \makecell{98.86} & \makecell{98.30} \\
    \makecell{GPT-2 256 MGTokenizer CustomTokenizer} & \makecell{99.12} & \makecell{98.88} & \makecell{100} & \makecell{99.44} \\
    \makecell{XLNet 256 MGTokenizer CustomTokenizer} & \makecell{94.69} & \makecell{97.67} & \makecell{95.45} & \makecell{96.55} \\
    \bottomrule
\end{tabular}
\end{table}

\section{Discussion}
\label{sec:discussfuture}
The main objective of this research was to evaluate the Transformer models, using varied feature engineering approaches, on their ability to detect novel ransomware. Consequently, any function present in the test samples that also appeared in the training data was removed from the test samples. As a result, for accurate classification, the model was forced to determine which combinations of ASM instructions and their contexts were malicious and which were benign.

In Experiment A the SVM hyperplane intercepted the x axis within the range of data, this is because there was only one benign sample with 100-1000 functions. The dataset for Experiment B was re-balanced to include benign samples with 100-1000 and 1000+ instructions in the test set which resulted in a more reasonable separation between classes. 

After de-duplication and filtering several test samples were left with fewer than 10 functions. Specifically, there were 2 Lockbit samples where after that removal of previously seen functions only 3 functions remained. The only 2 models that correctly classified those samples as malicious were DistilBERT 256 CustomTokenizer and GPT-2 256 CustomTokenizer. These two models also demonstrated the highest performance overall, where only one sample was misclassified. The benign sample \url{262aa82cfd43744433d44228d12c28b322970afd382587ad82f6bb07b903a501}, named googleupdatebroker.exe, was flagged as malicious by both of these models and 5 out of the 6 models in total. Consequently, we investigated it further. This file interacts with numerous registry keys and imports several suspicious APIs, including: \texttt{TerminateProcess}, \texttt{IsDebuggerPresent}, \texttt{UnhandledExceptionFilter}, \texttt{GetCommandLineW}, \texttt{GetProcAddress}. It also employs the \texttt{SetUnhandledExceptionFilter} function, which is commonly used as an anti-debugging technique, and queries the system time. These characteristics, combined with the file's overall behavior, raise significant concerns about its potential malicious intent. Although it is labeled as benign, the observed features suggest that it might not be a false positive.


A comparison to other approaches is shown in Table \ref{tab:performance_comps}. Pulse outperforms the others in several key areas. Our method uniquely detects truly new and previously unseen samples and achieves the highest accuracy. There is often little difference between ransomware variants, typically using slight variations on the file enumeration and encryption algorithms. Therefore, if a model has been trained on a variant of a sample present in the test data, it should be relatively easy for the model to correctly classify the test sample. Uniquely, our approach removes any familiar functions from the test samples, forcing the model to infer malicious behavior based solely on context and new ASM instruction combinations.

\begin{table}[t]
\centering
\footnotesize
\captionsetup{justification=centering}
\caption{A comparison of the Pulse framework with state-of-the-art ransomware detection \\ \textbf{Note:} b=benign, r=ransomware}
\label{tab:performance_comps}
\begin{tabular}{ccccc}
    \toprule
    \makecell{Paper} & \makecell{Dataset} & \makecell{Features} & \makecell{Model} & \makecell{Accuracy \%}\\
    \midrule
    \makecell{\cite{deng2024}} & \makecell{27,118 B\\35,367 R} & \makecell{Static PE \\Header} & \makecell{Double Deep Q\\Learning Network}& \makecell{97.9} \\\\
    \makecell{\cite{manavi2022}} & \makecell{1,000 B\\879 R} & \makecell{Static PE \\Header} & \makecell{Random Forest}& \makecell{96.80} \\\\
    \makecell{\cite{zahoora2022}} & \makecell{942 B\\582 R} & \makecell{Sandbox} & \makecell{Auto Encoder\\Ensemble Classifier}& \makecell{92.8} \\\\
    \makecell{\cite{Sharmeen2020}} & \makecell{754 B\\483 R} & \makecell{Sandbox} & \makecell{Deep Learning\\Fast ICA}& \makecell{95.96} \\\\
    \makecell{\cite{zhang2024ransomware}} & \makecell{-} & \makecell{Network Traffic\\File System} & \makecell{Clustering\\Deep Learning}& \makecell{95.00} \\\\
    \makecell{\cite{ayub2024}} & \makecell{101 B\\215 R} & \makecell{Sandbox} & \makecell{Random Forest}& \makecell{97.67} \\\\
    \makecell{\cite{zahoora2022pareto}} & \makecell{942 B\\582 R} & \makecell{Sandbox} & \makecell{Cost-Sensitive\\Pareto Ensemble}& \makecell{93.00} \\\\
    \makecell{Pulse\\(our work)} & \makecell{308 B\\526 R} & \makecell{Peekaboo DBI} & \makecell{Transformers}& \makecell{99.12} \\
    \bottomrule
\end{tabular}
\end{table}

\section{Conclusion and future research}
\label{sec:conclusion} 
Various Transformer models were trained with several feature engineering approaches using Peekaboo DBI data. The test data consisted of several ransomware families and benign samples. One of the main objectives of this research was novel ransomware detection, therefore any function present in the test samples that also appeared in the training data was removed from the test samples. Consequently, for the model to make the correct classification it had to infer what ASM instruction combinations and context were malicious and which were benign. The function classification layer predicted the number of benign functions and malicious functions for each sample. The output layer performed the final classification for the test samples by utilizing a Support Vector Machine (SVM) hyperplane, which acts as the decision boundary to delineate between benign and malicious samples. Our experimental results demonstrate the effectiveness and transferability of Pulse, which is robust to detecting previously unseen malicious functions. Pulse outperforms the previous state-of-the-art approaches for novel ransomware detection. Unlike other models that may use familiar functions across train and test samples, our approach removes these, forcing the model to detect malicious behavior based solely on context and new ASM instruction combinations. To the best of our knowledge this is the first research to use ASM language with Transformer models to detect never before seen malicious functions and samples.

In future research we will focus on extending the techniques presented in this research to the other types of malware available in the Peekaboo dataset, including Worms, Spyware, Trojans, Botnets, APT and post exploitation tools, each with distinct objectives and functionality \cite{gaber2024}. A critical aspect to the highly accurate models presented in Experiment B is the high quality data captured by Peekaboo. In addition to the semantic information captured in our feature engineering approach, the Peekaboo data also incorporates temporal information and complete network traffic captures and this data could be leveraged in future work. In terms of a real world application of the techniques presented in this research, a limitation is the time it takes Peekaboo to analyse a sample and extract the ASM instructions, in future work we will determine how much information is required by the Transformer models to make an accurate classification.

\section{Data Availability Statement}
\label{sec:Data Availability Statement}
The fine tuned Transformer models and associated scripts are available in the Peekaboo Transformer Models repository \cite{gabertpro2024}.

\section{Author Contributions Statement}
\label{sec:Author Contributions Statement}
Matthew Gaber conceived of the presented idea, carried out the experiments, wrote the main manuscript text and prepared the figures. Mohiuddin Ahmed and Helge Janicke helped supervise the project. All authors discussed the results and contributed to the manuscript.

 \bibliographystyle{elsarticle-num} 
 \bibliography{cas-refs}

\begin{thebibliography}{10}
\expandafter\ifx\csname url\endcsname\relax
  \def\url#1{\texttt{#1}}\fi
\expandafter\ifx\csname urlprefix\endcsname\relax\def\urlprefix{URL }\fi
\expandafter\ifx\csname href\endcsname\relax
  \def\href#1#2{#2} \def\path#1{#1}\fi

\bibitem{gabercsur2024}
M.~G. Gaber, M.~Ahmed, H.~Janicke, \href{https://doi.org/10.1145/3638552}{Malware detection with artificial intelligence: A systematic literature review}, ACM Comput. Surv. 56~(6) (jan 2024).
\newblock \href {https://doi.org/10.1145/3638552} {\path{doi:10.1145/3638552}}.
\newline\urlprefix\url{https://doi.org/10.1145/3638552}

\bibitem{xiao2020}
G.~Xiao, J.~Li, Y.~Chen, K.~Li, Malfcs: An effective malware classification framework with automated feature extraction based on deep convolutional neural networks, Journal of Parallel and Distributed Computing 141 (2020) 49--58.
\newblock \href {https://doi.org/10.1016/j.jpdc.2020.03.012} {\path{doi:10.1016/j.jpdc.2020.03.012}}.

\bibitem{gaber2024}
M.~Gaber, M.~Ahmed, H.~Janicke, Defeating evasive malware with peekaboo: Extracting authentic malware behavior with dynamic binary instrumentation, ResearchSquare Preprint (2024).
\newblock \href {https://doi.org/https://doi.org/10.21203/rs.3.rs-4279929/v1} {\path{doi:https://doi.org/10.21203/rs.3.rs-4279929/v1}}.

\bibitem{sophos2024}
Sophos, \href{https://assets.sophos.com/X24WTUEQ/at/9brgj5n44hqvgsp5f5bqcps/sophos-state-of-ransomware-2024-wp.pdf}{The state of ransomware 2024} (2024).
\newline\urlprefix\url{https://assets.sophos.com/X24WTUEQ/at/9brgj5n44hqvgsp5f5bqcps/sophos-state-of-ransomware-2024-wp.pdf}

\bibitem{ibmsecurity2023}
IBMSecurity, \href{https://www.ibm.com/downloads/cas/E3G5JMBP}{Cost of a data breach report 2023} (2023).
\newline\urlprefix\url{https://www.ibm.com/downloads/cas/E3G5JMBP}

\bibitem{kerns2022}
Q.~Kerns, B.~Payne, T.~Abegaz, Double-extortion ransomware: A technical analysis of maze ransomware, in: K.~Arai (Ed.), Proceedings of the Future Technologies Conference (FTC) 2021, Volume 3, Springer International Publishing, Cham, 2022, pp. 82--94.

\bibitem{payne2021}
B.~Payne, E.~Mienie, Multiple-extortion ransomware: The case for active cyber threat intelligence, in: ECCWS 2021 20th European Conference on Cyber Warfare and Security, Vol.~6, Academic Conferences Inter Ltd, 2021, pp. 331--336.

\bibitem{meurs2022}
T.~Meurs, M.~Junger, E.~Tews, A.~Abhishta, Ransomware: How attacker’s effort, victim characteristics and context influence ransom requested, payment and financial loss, in: 2022 APWG Symposium on Electronic Crime Research (eCrime), 2022, pp. 1--13.
\newblock \href {https://doi.org/10.1109/eCrime57793.2022.10142138} {\path{doi:10.1109/eCrime57793.2022.10142138}}.

\bibitem{ahn2022}
S.~Ahn, S.~Ahn, H.~Koo, Y.~Paek, \href{http://dx.doi.org/10.1145/3564625.3567975}{Practical binary code similarity detection with bert-based transferable similarity learning}, Proceedings of the 38th Annual Computer Security Applications Conference (2022) 361\href {https://doi.org/10.1145/3564625.3567975} {\path{doi:10.1145/3564625.3567975}}.
\newline\urlprefix\url{http://dx.doi.org/10.1145/3564625.3567975}

\bibitem{aurangzeb2021}
S.~Aurangzeb, R.~N.~B. Rais, M.~Aleem, M.~A. Islam, M.~A. Iqbal, On the classification of microsoft-windows ransomware using hardware profile, PeerJ. Computer Science 7 (2021) e361.
\newblock \href {https://doi.org/10.7717/peerj-cs.361} {\path{doi:10.7717/peerj-cs.361}}.

\bibitem{HIRANO2022}
M.~Hirano, R.~Hodota, R.~Kobayashi, \href{https://www.sciencedirect.com/science/article/pii/S2666281721002390}{Ransap: An open dataset of ransomware storage access patterns for training machine learning models}, Forensic Science International: Digital Investigation 40 (2022) 301314.
\newblock \href {https://doi.org/https://doi.org/10.1016/j.fsidi.2021.301314} {\path{doi:https://doi.org/10.1016/j.fsidi.2021.301314}}.
\newline\urlprefix\url{https://www.sciencedirect.com/science/article/pii/S2666281721002390}

\bibitem{khan2020}
F.~Khan, C.~Ncube, L.~K. Ramasamy, S.~Kadry, Y.~Nam, A digital dna sequencing engine for ransomware detection using machine learning, IEEE Access 8 (2020) 119710--119719.
\newblock \href {https://doi.org/10.1109/ACCESS.2020.3003785} {\path{doi:10.1109/ACCESS.2020.3003785}}.

\bibitem{CARLIN2019}
D.~Carlin, P.~O’Kane, S.~Sezer, \href{https://www.sciencedirect.com/science/article/pii/S0167404819300082}{A cost analysis of machine learning using dynamic runtime opcodes for malware detection}, Computers \& Security 85 (2019) 138--155.
\newblock \href {https://doi.org/https://doi.org/10.1016/j.cose.2019.04.018} {\path{doi:https://doi.org/10.1016/j.cose.2019.04.018}}.
\newline\urlprefix\url{https://www.sciencedirect.com/science/article/pii/S0167404819300082}

\bibitem{gibert2021}
D.~Gibert, C.~Mateu, J.~Planes, J.~Marques-Silva, Auditing static machine learning anti-malware tools against metamorphic attacks, Computers \& Security 102 (2021).
\newblock \href {https://doi.org/10.1016/j.cose.2020.102159} {\path{doi:10.1016/j.cose.2020.102159}}.

\bibitem{ye2017}
Y.~Ye, T.~Li, D.~Adjeroh, S.~S. Iyengar, A survey on malware detection using data mining techniques, ACM Computing Surveys (CSUR) 50~(3) (2017) 1--40.
\newblock \href {https://doi.org/10.1145/3073559} {\path{doi:10.1145/3073559}}.

\bibitem{vonderassen2023}
J.~von~der Assen, A.~H. Celdrán, J.~Luechinger, P.~M.~S. Sánchez, G.~Bovet, G.~M. Pérez, B.~Stiller, \href{https://arxiv.org/abs/2306.15559}{Ransomai: Ai-powered ransomware for stealthy encryption} (2023).
\newblock \href {http://arxiv.org/abs/2306.15559} {\path{arXiv:2306.15559}}.
\newline\urlprefix\url{https://arxiv.org/abs/2306.15559}

\bibitem{Kajiwara2021}
Y.~Kajiwara, J.~Zheng, K.~Mouri, Performance comparison of training datasets for system call-based malware detection with thread information, IEICE Transactions on Information and Systems E104D~(12) (2021) 2173--2183.

\bibitem{ucci2019}
D.~Ucci, L.~Aniello, R.~Baldoni, Survey of machine learning techniques for malware analysis, Computers \& Security 81 (2019) 123.
\newblock \href {https://doi.org/10.1016/j.cose.2018.11.001} {\path{doi:10.1016/j.cose.2018.11.001}}.

\bibitem{galloro2022}
N.~Galloro, M.~Polino, M.~Carminati, A.~Continella, S.~Zanero, A systematical and longitudinal study of evasive behaviors in windows malware, Computers \& security 113 (2022).

\bibitem{kim2022}
M.~Kim, H.~Cho, J.~H. Yi, Large-scale analysis on anti-analysis techniques in real-world malware, IEEE access 10 (2022) 75802--75815.

\bibitem{maffia2021}
L.~Maffia, D.~Nisi, P.~Kotzias, G.~Lagorio, S.~Aonzo, D.~Balzarotti, \href{https://arxiv.org/abs/2112.11289}{Longitudinal study of the prevalence of malware evasive techniques}, ArXiv (2021).
\newline\urlprefix\url{https://arxiv.org/abs/2112.11289}

\bibitem{park2019}
J.~Park, Y.~H. Jang, S.~Hong, Y.~Park, Automatic detection and bypassing of anti-debugging techniques for microsoft windows environments (2019).
\newblock \href {https://doi.org/10.4316/AECE.2019.02003} {\path{doi:10.4316/AECE.2019.02003}}.

\bibitem{nunes2022}
M.~Nunes, P.~Burnap, P.~Reinecke, K.~Lloyd, Bane or boon: Measuring the effect of evasive malware on system call classifiers, Journal of Information Security and Applications 67 (2022).
\newblock \href {https://doi.org/10.1016/j.jisa.2022.103202} {\path{doi:10.1016/j.jisa.2022.103202}}.

\bibitem{gaberro2024}
M.~Gaber, M.~Ahmed, H.~Janicke, \href{https://ro.ecu.edu.au/datasets/138/}{Peekaboo} (2024).
\newblock \href {https://doi.org/https://doi.org/10.25958/85p1-4w32} {\path{doi:https://doi.org/10.25958/85p1-4w32}}.
\newline\urlprefix\url{https://ro.ecu.edu.au/datasets/138/}

\bibitem{koo2021}
H.~Koo, S.~Park, D.~Choi, T.~Kim, Semantic-aware Binary Code Representation with BERT, Cornell University Library, arXiv.org, Ithaca, 2021.

\bibitem{gabertpro2024}
M.~Gaber, M.~Ahmed, H.~Janicke, \href{https://ro.ecu.edu.au/datasets/142/}{Peekaboo transformer models} (2024).
\newblock \href {https://doi.org/https://doi.org/10.25958/z82g-1e40} {\path{doi:https://doi.org/10.25958/z82g-1e40}}.
\newline\urlprefix\url{https://ro.ecu.edu.au/datasets/142/}

\bibitem{devlin2019}
J.~Devlin, M.-W. Chang, K.~Lee, K.~Toutanova, \href{https://aclanthology.org/N19-1423}{{BERT}: Pre-training of deep bidirectional transformers for language understanding}, in: J.~Burstein, C.~Doran, T.~Solorio (Eds.), Proceedings of the 2019 Conference of the North {A}merican Chapter of the Association for Computational Linguistics: Human Language Technologies, Volume 1 (Long and Short Papers), Association for Computational Linguistics, Minneapolis, Minnesota, 2019, pp. 4171--4186.
\newblock \href {https://doi.org/10.18653/v1/N19-1423} {\path{doi:10.18653/v1/N19-1423}}.
\newline\urlprefix\url{https://aclanthology.org/N19-1423}

\bibitem{vaswani2017}
A.~Vaswani, N.~Shazeer, N.~Parmar, J.~Uszkoreit, L.~Jones, A.~N. Gomez, L.~Kaiser, I.~Polosukhin, Attention is all you need, Advances in neural information processing systems 30 (2017).

\bibitem{sanh2020}
V.~Sanh, L.~Debut, J.~Chaumond, T.~Wolf, DistilBERT, a distilled version of BERT: smaller, faster, cheaper and lighter, Cornell University Library, arXiv.org, Ithaca, 2020.

\bibitem{liu2019}
Y.~Liu, M.~Ott, N.~Goyal, J.~Du, M.~Joshi, D.~Chen, O.~Levy, M.~Lewis, L.~Zettlemoyer, V.~Stoyanov, RoBERTa: A Robustly Optimized BERT Pretraining Approach, Cornell University Library, arXiv.org, Ithaca, 2019.

\bibitem{huggingface2024}
HuggingFace, \href{https://huggingface.co/distilbert/distilroberta-base}{distilroberta-base} (2024).
\newline\urlprefix\url{https://huggingface.co/distilbert/distilroberta-base}

\bibitem{radford2019}
A.~Radford, J.~Wu, R.~Child, D.~Luan, D.~Amodei, I.~Sutskever, \href{https://cdn.openai.com/better-language-models/language_models_are_unsupervised_multitask_learners.pdf}{Language models are unsupervised multitask learners} (2019).
\newline\urlprefix\url{https://cdn.openai.com/better-language-models/language_models_are_unsupervised_multitask_learners.pdf}

\bibitem{yang2019}
Z.~Yang, Z.~Dai, Y.~Yang, J.~Carbonell, R.~Salakhutdinov, Q.~V. Le, XLNet: generalized autoregressive pretraining for language understanding, Curran Associates Inc., Red Hook, NY, USA, 2019.

\bibitem{thurner2015}
S.~Thurner, R.~Hanel, B.~Liu, B.~Corominas-Murtra, \href{http://dx.doi.org/10.1098/rsif.2015.0330}{Understanding zipf's law of word frequencies through sample-space collapse in sentence formation}, Journal of The Royal Society Interface 12~(108) (2015).
\newblock \href {https://doi.org/10.1098/rsif.2015.0330} {\path{doi:10.1098/rsif.2015.0330}}.
\newline\urlprefix\url{http://dx.doi.org/10.1098/rsif.2015.0330}

\bibitem{li2021}
X.~Li, Y.~Qu, H.~Yin, \href{http://dx.doi.org/10.1145/3460120.3484587}{Palmtree learning an assembly language model for instruction embedding} (2021).
\newblock \href {https://doi.org/10.1145/3460120.3484587} {\path{doi:10.1145/3460120.3484587}}.
\newline\urlprefix\url{http://dx.doi.org/10.1145/3460120.3484587}

\bibitem{demirkiran2022}
F.~Demirkıran, A.~Çayır, U.~Ünal, H.~Dağ, \href{http://dx.doi.org/10.1016/j.cose.2022.102846}{An ensemble of pre-trained transformer models for imbalanced multiclass malware classification}, Computers \& Security 121 (2022).
\newblock \href {https://doi.org/10.1016/j.cose.2022.102846} {\path{doi:10.1016/j.cose.2022.102846}}.
\newline\urlprefix\url{http://dx.doi.org/10.1016/j.cose.2022.102846}

\bibitem{rahali2021}
A.~Rahali, M.~A. Akhloufi, Malbert: Malware detection using bidirectional encoder representations from transformers, in: 2021 IEEE International Conference on Systems, Man, and Cybernetics (SMC), 2021, pp. 3226--3231.
\newblock \href {https://doi.org/10.1109/SMC52423.2021.9659287} {\path{doi:10.1109/SMC52423.2021.9659287}}.

\bibitem{maniriho2024}
P.~Maniriho, A.~N. Mahmood, M.~J.~M. Chowdhury, \href{https://arxiv.org/abs/2407.13355}{Earlymaldetect: A novel approach for early windows malware detection based on sequences of api calls} (2024).
\newblock \href {http://arxiv.org/abs/2407.13355} {\path{arXiv:2407.13355}}.
\newline\urlprefix\url{https://arxiv.org/abs/2407.13355}

\bibitem{LIU2024SeMalBERT}
J.~Liu, Y.~Zhao, Y.~Feng, Y.~Hu, X.~Ma, \href{https://www.sciencedirect.com/science/article/pii/S2214212623002740}{Semalbert: Semantic-based malware detection with bidirectional encoder representations from transformers}, Journal of Information Security and Applications 80 (2024) 103690.
\newblock \href {https://doi.org/https://doi.org/10.1016/j.jisa.2023.103690} {\path{doi:https://doi.org/10.1016/j.jisa.2023.103690}}.
\newline\urlprefix\url{https://www.sciencedirect.com/science/article/pii/S2214212623002740}

\bibitem{lu2024malsightexploringmalicioussource}
H.~Lu, H.~Peng, G.~Nan, J.~Cui, C.~Wang, W.~Jin, \href{https://arxiv.org/abs/2406.18379}{Malsight: Exploring malicious source code and benign pseudocode for iterative binary malware summarization} (2024).
\newblock \href {http://arxiv.org/abs/2406.18379} {\path{arXiv:2406.18379}}.
\newline\urlprefix\url{https://arxiv.org/abs/2406.18379}

\bibitem{googlecolab}
{Google}, \href{https://colab.research.google.com/}{Google colaboratory}, accessed: 2024-03-11 (2024).
\newline\urlprefix\url{https://colab.research.google.com/}

\bibitem{deng2024}
X.~Deng, M.~Cen, M.~Jiang, M.~Lu, \href{https://doi.org/10.1007/s10586-023-04043-5}{Ransomware early detection using deep reinforcement learning on portable executable header}, Cluster Computing~(27) (may 2024).
\newblock \href {https://doi.org/10.1007/s10586-023-04043-5} {\path{doi:10.1007/s10586-023-04043-5}}.
\newline\urlprefix\url{https://doi.org/10.1007/s10586-023-04043-5}

\bibitem{manavi2022}
F.~Manavi, A.~Hamzeh, \href{https://doi.org/10.1007/s11416-021-00414-x}{A novel approach for ransomware detection based on pe header using graph embedding}, Journal of Computer Virology and Hacking Techniques~(18) (december 2022).
\newblock \href {https://doi.org/10.1007/s11416-021-00414-x} {\path{doi:10.1007/s11416-021-00414-x}}.
\newline\urlprefix\url{https://doi.org/10.1007/s11416-021-00414-x}

\bibitem{zahoora2022}
U.~Zahoora, M.~Rajarajan, Z.~Pan, A.~Khan, \href{https://doi.org/10.1007/s10489-022-03244-6}{Ransomware attack detection using deep contractive autoencoder and voting based ensemble classifier}, Applied Intelligence~(52) (september 2022).
\newblock \href {https://doi.org//10.1007/s10489-022-03244-6} {\path{doi:/10.1007/s10489-022-03244-6}}.
\newline\urlprefix\url{https://doi.org/10.1007/s10489-022-03244-6}

\bibitem{Sharmeen2020}
S.~Sharmeen, Y.~A. Ahmed, S.~Huda, B.~S. Kocer, M.~M. Hassan, \href{https://dro.deakin.edu.au/articles/journal_contribution/Avoiding_future_digital_extortion_through_robust_protection_against_ransomware_threats_using_deep_learning_based_adaptive_approaches/20716468}{Avoiding future digital extortion through robust protection against ransomware threats using deep learning based adaptive approaches}, IEEE Access Special Section on Deep Learning: Security and Forensics Research Advances and Challenges (1 2020).
\newblock \href {https://doi.org/10.1109/ACCESS.2020.2970466} {\path{doi:10.1109/ACCESS.2020.2970466}}.
\newline\urlprefix\url{https://dro.deakin.edu.au/articles/journal_contribution/Avoiding_future_digital_extortion_through_robust_protection_against_ransomware_threats_using_deep_learning_based_adaptive_approaches/20716468}

\bibitem{zhang2024ransomware}
R.~Zhang, Y.~Liu, Ransomware detection with a 2-tier machine learning approach using a novel clustering algorithm, ResearchSquare (2024).

\bibitem{ayub2024}
A.~Ayub, A.~Siraj, B.~Filar, M.~Gupta, \href{https://doi.org/10.1007/s10207-023-00758-z}{Rwarmor: a static-informed dynamic analysis approach for early detection of cryptographic windows ransomware}, International Journal of Information Security~(53) (february 2024).
\newblock \href {https://doi.org/10.1007/s10207-023-00758-z} {\path{doi:10.1007/s10207-023-00758-z}}.
\newline\urlprefix\url{https://doi.org/10.1007/s10207-023-00758-z}

\bibitem{zahoora2022pareto}
U.~Zahoora, A.~Khan, M.~Rajarajan, S.~Hussain~Khan, M.~Asam, T.~Jamal, \href{https://doi.org/10.1038/s41598-022-19443-7}{Ransomware detection using deep learning based unsupervised feature extraction and a cost sensitive pareto ensemble classifier}, Scientific Reports~(12) (september 2022).
\newblock \href {https://doi.org/10.1038/s41598-022-19443-7} {\path{doi:10.1038/s41598-022-19443-7}}.
\newline\urlprefix\url{https://doi.org/10.1038/s41598-022-19443-7}

\end{thebibliography}





\end{document}